\documentclass[conference]{IEEEtran}
\IEEEoverridecommandlockouts
\usepackage{cite}
\usepackage{amsmath,amssymb,amsfonts}
\usepackage{graphicx}
\usepackage{subfig}
\usepackage{textcomp}
\usepackage{xcolor}
\usepackage{lipsum}
\usepackage{mathtools}
\def\BibTeX{{\rm B\kern-.05em{\sc i\kern-.025em b}\kern-.08em
    T\kern-.1667em\lower.7ex\hbox{E}\kern-.125emX}}
    
\newtheorem{definition}{Definition}
\usepackage[bookmarks=false]{hyperref}
\usepackage[capitalise]{cleveref}

\Crefname{equation}{Eq.}{Eqs.}
\Crefname{figure}{Fig.}{Figs.}
\Crefname{tabular}{Tab.}{Tabs.}

\newcommand{\optch}{OptChain}
\newcommand{\crsh}{cross-TX}

\usepackage{algorithm}
\usepackage{algpseudocode}

\usepackage{booktabs}
\usepackage[normalem]{ulem}
\usepackage{array}
\renewcommand{\baselinestretch}{0.92}    
\begin{document}

\title{\optch: Optimal Transactions Placement for Scalable Blockchain Sharding\\
}

\author{\IEEEauthorblockN{Lan N. Nguyen, Truc D. T. Nguyen 
\thanks{The first two authors contribute equally to this paper.}}\IEEEauthorblockA{CISE Department\\ University of Florida\\ Gainesville, FL, 32611\\ Email: \{lan.nguyen,truc.nguyen\}@ufl.edu}
  \and
  \IEEEauthorblockN{Thang N. Dinh}
  \IEEEauthorblockA{CS Department \\ Virginia Commonwealth University \\ Richmond, VA, 23284\\Email: tndinh@vcu.edu} \and
  \IEEEauthorblockN{My T. Thai}\IEEEauthorblockA{CISE Department\\ University of Florida\\ Gainesville, FL, 32611\\ Email: mythai@cise.ufl.edu}
}

\maketitle

\begin{abstract}
A major challenge in blockchain sharding protocols is that more than 95\%  transactions are cross-shard. Not only those cross-shard transactions degrade the system throughput but also double the confirmation time, and exhaust an already scarce network bandwidth. Are cross-shard transactions imminent for sharding schemes?
In this paper, we propose a new sharding paradigm, called \optch, in which cross-shard transactions are minimized, resulting in almost twice faster confirmation time and throughput. By treating transactions as a stream of nodes in an online graph,  OptChain utilizes a lightweight and on-the-fly transaction placement method to group both related and soon-related transactions into the same shards. At the same time, OptChain maintains a temporal balance among shards to guarantee the high parallelism. Our comprehensive and large-scale simulation using Oversim P2P library confirms a significant boost in performance with up to 10 folds reduction in cross-shard transactions, more than twice reduction in confirmation time, and 50\% increase in throughput. When combined with Omniledger sharding protocol, OptChain delivers a 6000 transactions per second throughput with 10.5s confirmation time.
\end{abstract}



\section{Introduction}
Blockchain has emerged as a disruptive and transformational technology, with great potential and benefits, offering a promising new decentralized economy without the risk of  single point of failures, monopoly, or censorship \cite{swan2015blockchain}. Examples of these systems are ranging from the cryptocurrency such as Bitcoin \cite{nakamoto2008bitcoin} and Ethereum \cite{ethereum}, to other infrastructures and application domains such as the Internet-of-Things \cite{christidis2016blockchains,huckle2016internet} and Digitial Health \cite{yue2016healthcare,azaria2016medrec}. Unfortunately, the performance level of existing decentralized systems based on the Blockchain technology is too low to realize that vision, e.g., 7 transactions per second (tps) and up to 60 minutes confirmation time for Bitcoin and 10 tps/12 minutes for Ethereum. 




To this end, \emph{blockchain sharding}, which splits the Blockchain into multiple disjoint parts, called shards, each maintained by a subgroup of nodes, has been proposed as a prominent solution for Blockchain scaling. Since each node only needs to communicate with a few (hundreds)  nodes from the same shard to maintain a small chunk of blockchain, sharding reduces substantially the storage, computation, and communication costs. This is different from legacy blockchain systems, e.g., Bitcoin and Ethereum, in which all nodes need to communicate to maintain the same copy of blockchain, thus, the performance is limited by the nodes average processing capabilities. The latest sharding approaches such as Omniledger \cite{kokoris2017omniledger} and Rapidchain \cite{zamanirapidchain} can handle thousands of transactions per second with confirmation time about a few dozens of seconds. 

All existing sharding approaches, however,  face the same challenge of handling \emph{cross-shard} transactions, which involve data from more than one shards. To prevent the double-spending problem \cite{karame2015misbehavior}, all shards that involve a cross-shard transaction (tx) need to execute multiple-phase protocols to confirm the tx. This can multiple fold increase both the latency and the effort to confirm the tx, comparing to the case when the tx need to be processed by only one shard. In turns, extra efforts in confirming txs may lead to higher tx fees. To make it even worse, more than 95\% of the transactions are cross-shard \cite{kokoris2017omniledger, zamanirapidchain}. Previous sharding approaches often place txs into shards randomly to balance the load among the shards. It is natural to ask ``is there a smart transactions placement strategy that reduces the cross-shard txs, making sharding even faster?''

In this paper, we propose \optch, a sharding-agnostic framework that boosts the performance of existing (and future) sharding approaches via optimizing the placement of txs into shards. \optch~learns the pattern from the past txs to decide the shard-location for incoming txs based on 1) whether such placement reduces the cross-shard txs and 2) load balance among the shards. Specifically, \optch~works on top an unexplored graph construction for transaction networks in UTXO model \cite{nakamoto2008bitcoin}, termed \emph{Transactions as Nodes} (TaN), and introduces a new score, termed \emph{Temporal Fitness} to assess the suitability of placing an incoming transaction to each shard.

The \emph{TaN network} is constructed by abstracting each transaction as a node and there is a directed edge $(u, v)$ if tx $u$ uses tx $v$ as an input. This TaN is an online directed acyclic graph (DAG), in which nodes arriving one by one. This construction is different from existing abstraction of transaction networks in which transactions are abstracted as edges among the nodes made of addresses \cite{kondor2014rich}.

The \emph{Temporal Fitness} score is composed by two component scores \emph{Transaction-to-Shard} (T2S) and  \emph{Latency-to-Shard} (L2S). The T2S scores between a transaction $u$ and a shard $S_i$, measures the probability that a random walk from a node/tx $u$ ends up at some node in $S_i$ (see section IV.B), i.e., hence, how likely the transaction should be placed into the shard without causing further crosshard txs. The L2S score estimates the processing delay when placing the transaction to a shard. Importantly, the protocol to estimate the two scores is lightweight and is executed at the users side.



\textbf{Practicality.} Our solution $\optch$ can be implemented with simple modification in user-side software, e.g., wallet. That is $\optch$ does not interfere with the core consensus protocols and, hence, can be integrated into almost all sharding approaches. Specifically, as computing the T2S score only requires the information on the input txs, it can be done efficiently at the user side by modifying the existing Simple Payment Verification protocol \cite{nakamoto2008bitcoin}, i.e., users do not need to download the complete transaction history. Based on the latencies observed from the shard, the wallet software at the user side can use \optch~ to make the decision on which shard he/she wants the tx to be processed.

To validate our approach, we measure the performance of an enhanced version of  OmniLedger \cite{kokoris2017omniledger} with our \optch~approach on existing Bitcoin transactions. The results indicate that \optch~can effectively reduce the cross-shard txs up to 10 folds, cut the txs confirmation time by 93\%, and at the same time, increase the throughput by 50\%. While we only test \optch~with Omniledger, we predict a similar level of improvement in performance when combining \optch~with other sharding protocols such as Rapidchain.

\textbf{Our contribution.} We summarize our contribution as follows.
\begin{itemize}
\item We introduce a new way of sharding txs, reducing cross-shard txs via an optimal placement of txs into shards. This simple idea effectively reduces the cross-shard txs and boosts the performance comparing to the random placement in existing sharding approaches.

\item We investigate a new abstraction of transactions network (TaN) in which transactions are abstracted as nodes rather than edges among addresses in previous studies. This new abstraction results in an online DAG that can provide new ways to analyze the transaction stream in UTXO-based ledgers.  We provide various charateristics of this TaN on Bitcoin transactions consisting of 298,325,121 nodes and 696,860,716 edges.

\item We introduce a novel algorithm, called \optch, that analyze the stream of txs in TaN network to make the optimal placement of txs into shards. \optch~is simple, lightweight, and can be easily implemented into existing wallet software.

\item Our comprehensive experiments with an enhanced Omniledger protocol on real Bitcoin transactions affirm the significant benefit of our approach in cutting down the confirmation time and boosting throughput.

\end{itemize}

{\bf Organization.} The rest of this paper is structured as follows. We summarize the related work in Section \ref{sec:related}. Section \ref{sec:blockchain-sharding} discusses the basis of handling cross-shard transactions in blockchain sharding and provides some observations as well as primary goals of the transactions placement strategies. In Section \ref{sec:optchain}, we investigate the TaN network and present our OptChain algorithm. The experimental design and results are presented in Section \ref{sec:experiment}. Finally, Section \ref{sec:conclusion} gives the concluding remarks.

\section{Related work}\label{sec:related}

\textbf{Blockchain sharding.}
Several  blockchain sharding  protocols  \cite{george2015centrally,luu2016secure, kokoris2017omniledger,zamanirapidchain,chainspace18,ethereum2}  have  been  proposed  to address the  scalability  issue  in  legacy  blockchain. In typical blockchain sharding, the  entire state of the blockchain is splitted into  partitions called \textit{shards} that contain their own independent piece of state and transaction history. The key idea is to parallelize the available computation power, dividing it into several smaller shards where each of them processes a disjoint set of transactions. 

Currently, most of existing sharding protocols are built on top of the UTXOs model. The most notable exception is Ethereum 2.0 \cite{ethereum2} which is the next development phase of the Ethereum blockchain \cite{ethereum},  employing the account model. Unlike the UTXO model, each transaction in the account model has only one input and one output.

The three core components of an existing sharding protocol are 1) how to (randomly) assign nodes into shards to form shard committees; 2) an intra-shard consensus protocol to execute by shard committees (often a BFT protocol \cite{kokoris2017omniledger, zamanirapidchain} but also can be a Nakamoto-like protocol \cite{Monoxide19}); and 3) an atomic protocol for cross-shard transactions. In this work, we focus on the third component in which we aim to optimize the placement of transactions, hence mitigate the negative impact of cross-shard transactions on the performance of existing sharding approaches \cite{kokoris2017omniledger, zamanirapidchain}.

\textbf{Transaction networks.} 
In our method, we abstract the relation between transactions under a graph representation. In previous work, Kondor et al. \cite{kondor2014rich} addressed the transactions network as a graph where each node represents a user address, each directed edge between two nodes is created if there is at least one transaction between the corresponding addresses. Our work is fundamentally different in which we abstract transactions as nodes while an edge between two nodes represents the behavior that one transaction spends an output of the other. By such representation, we model the problem of transaction sharding to be an online graph partitioning problem with temporal balancing, where nodes in a graph is divided into disjoint subsets and the objective is to identify which subset should contain a new arriving node.

\textbf{Online and Balanced DAG Partitioning.} 
Online graph partitioning has been addressed in the literature \cite{stanton2012streaming, abbas2018streaming}. Stanton et al. \cite{stanton2012streaming} and Abbas et al. \cite{abbas2018streaming} have proposed multiple natural, simple heuristics and compared their performance to Metis \cite{karypis1995metis}, a fast and offline method. The objective of their algorithms is to partition network vertices into almost equal disjoint sets while minimizing number of crossing edges (edge whose two endpoints are in two different set). However, these works are fundamentally different to our work, in which we would like to minimize number of cross-shard transactions rather than crossing edges. Furthermore, even their works eventually guarantee the balances between sets, we concern more on temporal balancing, in which we would like the number of vertices on each sets should be almost equal for any given of time.

\section{Cross-shard transactions and\\ Transactions Placement Strategies} \label{sec:blockchain-sharding}

In this section, we present an overview of state-of-the-art transaction sharding procedures. Although blockchain sharding itself could help boost the transactions confirmation process, nonetheless, the high amount of cross-shard transactions is one of the obstacles that hinder the system from getting better performance. Hence, we illustrate its negative impact which motivates us to devise the \optch{} to tolerate this issue.

\subsection{Atomic Commit Protocols for Cross-shard Transactions}

As both cross-shard protocols in \cite{kokoris2017omniledger} and \cite{zamanirapidchain} are built on top of ledgers using UTXO model, we begin with the presentation of this model. 

\textbf{Unspent Transaction outputs (UTXO) model.} The UTXO model was introduced in original Nakamoto protocol for Bitcoin \cite{nakamoto2008bitcoin}. In this model, transactions may have multiple inputs and outputs where an output is assigned with credits and locked to a user's address. This newly created transaction output is referred to as a UTXO. The UTXOs may then be used as inputs of another transaction, and after this transaction is committed to a block, those UTXOs will be marked as \textit{spent} and cannot be used again. For simplicity, we will consider a transaction $tx$ with two inputs $tx_{in}^1$, $tx_{in}^2$.

\textbf{Cross-shard Transactions.} Let $S^1_{in}, S^2_{in}, $ and $S_{out}$ denote the shards that contain $tx_{in}^1$, $tx_{in}^2$ and  $tx$, respectively. If all the shards $S^1_{in}, S^2_{in}, $ and $S_{out}$ are the same, we have a same-shard transaction, otherwise the transaction is cross-shard (\crsh).

If transactions are placed into shards randomly, it was shown  that the probability for a typical transaction having two inputs and one output to be a cross-shard transaction is about 94\% \cite{kokoris2017omniledger}, assuming 4 shards, and 99.98\%, assuming  16 shards.


\begin{figure}
	\centering
	\includegraphics[width=1\linewidth]{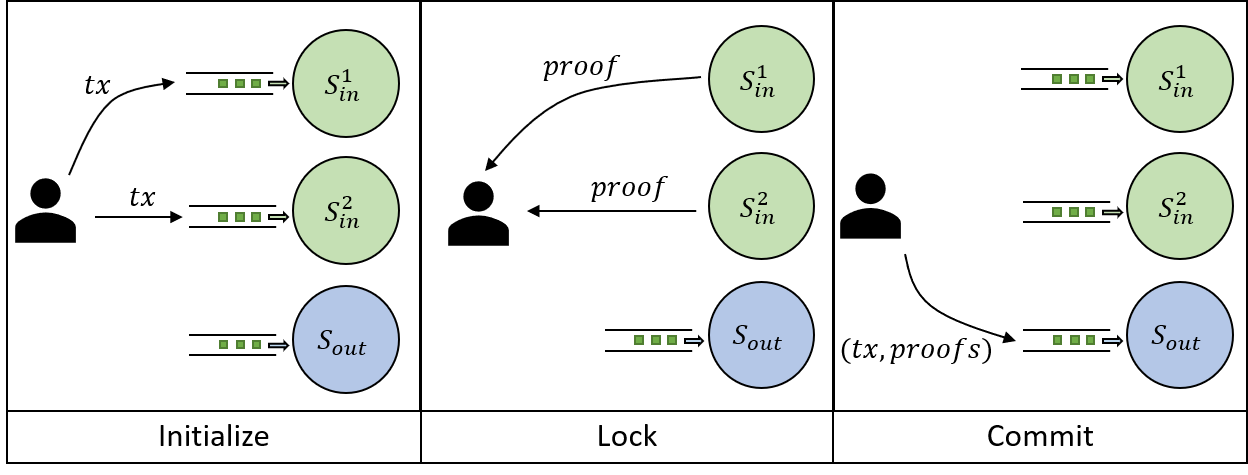}
	\caption{Handling cross-shard transactions in Omniledger\cite{kokoris2017omniledger}}
	\label{fig:cross-tx}
\end{figure}

\textbf{Committing Cross-TXs in OmniLedger} \cite{kokoris2017omniledger}. OmniLedger proposed a novel atomic protocol to commit \crsh s consistently among all shards.  The protocol locks all input transactions at the input shards before committing the output transaction(s) to output shard(s). 
\begin{enumerate}
    \item \textbf{Initialize}. A user creates a \crsh{} $tx$ whose inputs spend UTXOs, e.g., $tx_{in}^1$, $tx_{in}^2$, from some input shards, e.g., $S^1_{in}, S^2_{in}$. The client gossips the cross-TX and it eventually reaches all input shards.
    \item \textbf{Lock}. All input shards validate the transactions within his shard. If the transactions are valid, they are marked spent on the shard's ledger, and a \emph{proof-of-acceptance} is gossiped; otherwise a \emph{proof-of-rejection} is gossiped. 
    \item \textbf{Commit}. If all input shards gossip the proof-of-acceptance, the client can gossip an \emph{unlock-to-commit transactions} that eventually reach all output shards. In turn, each output shard validates the transaction and includes it to his ledger. However, if even one input shard issued a proof-of-rejection, then the transaction cannot be committed and has to abort. The client then can gossip an \emph{unlock-to-abort} message to reclaim the fund.
\end{enumerate}


\textbf{Committing Cross-TXs in  RapidChain} \cite{zamanirapidchain}. Rapidchain uses a ``yanking'' mechanism, in which the input transactions, e.g., $tx_{in}^1$, $tx_{in}^2$, will be first moved to from the input shards, e.g., $S^1_{in}, S^2_{in}$, to the output shard, e.g., $S_{out}$ via an inter-committee protocol. After all the input transactions are successfully ``yanked'' to the output shard, then the final transactions can be added to the ledger of the output shard.






\subsection{Performance Penalty for \crsh s.}
Comparing to same-shard transactions, \crsh{} incurs much higher confirmation time, communication and computation costs. 

\textbf{Longer confirmation time.} In Omniledger \cite{kokoris2017omniledger},  a \crsh{} will easily double the confirmation time of those in the same-shard transactions. For the same-shard transaction, the user only needs to submit the transaction to the shard and wait for confirmation. The confirmation time will only be the round time trip between the users and the shard pluses the time for the shard committee to agree on the transactions. A \crsh{} will incur extra time to confirm the input transactions as the input shards as well as another round time trip between the users and the shard committees. The same is applied for Rapidchain \crsh{} protocol as each \crsh{} incurs extra round-time trip among the committees as well as the waiting time for input transactions to agree on `yanking' transactions between shards.

\textbf{Extra communication and computation cost.} For a typical \crsh{} with 2 inputs and one output, the communication cost will triple that of a same-shard transaction as all the three shard committees and the user need to communicate to confirm the transaction. The same holds for the computation cost. 

Thus, assuming uniform time and cost to handle transactions in shards, each \crsh{} will \emph{double confirmation time} and \emph{triple the bandwith consumption and computation cost}. Therefore, if we can reduce the fraction of \crsh s to 20\%, we will cut the confirmation time by more than 40\%  and more than double the throughput.

\subsection{OptChain: A Transactions Placement Strategy}
\textbf{Random Placement}. In existing sharding approaches \cite{kokoris2017omniledger, zamanirapidchain}, transactions are placed randomly into shards. Often,  the hashed value of a transaction is used to determine which shards the transaction will be placed into. This will balance the amount of transactions per shard, however, cause almost all transactions to be \crsh s. For a typical transaction having two inputs and one output to be a cross-shard transaction is about 94\%, assuming 4 shards, and 99.98\%, assuming  16 shards \cite{kokoris2017omniledger}. To  be specific, the systems do not consider the relationship between transactions on sharding process, which makes a majority of transactions eventually become \crsh s.

\textbf{Smart Transaction Placement.} Ideally, the best method is to groups well-connected transactions into a same shard. By that way, we can minimize the number of verification steps (step 2) of \crsh to get proof-of-acceptance. Moreover, the current state of shards should also be considered. We would like to avoid situations where some shards are extremely busy (i.e huge number of transactions in queue to wait for verification) while some are idle. Intuitively, with a random selection, we can balance the number of transactions in each shards. However, some transactions could take more time to processed than the others. Therefore, in simulation, we observe that there are some certain moments the random selection will eventually cause extremely imbalance on queue sizes between shards.       

\textbf{OptChain.} Motivated by observations on the limits of OmniLedger and RapidChain, in this paper, we investigate an optimization problem in which users need to determine the best shard to submit their transactions in order to minimize \crsh s while guaranteeing the temporal balance, thus, shorten the confirmation time and boost the overall system throughput. Specifically, the ultimate goal of our smart transactions placement OptChain are:
\begin{enumerate}
     \item \textbf{Fast Confirmation Time}: As major txs are same-shard txs and load are distributed evenly, txs get confirmed in much shorter time.
    \item \textbf{High Throughput}: Same-shard txs require less time and communication to confirm, thus, the system throughput gets significantly boosted.
\end{enumerate}
The above goals are obtained via optimizing two indirect goals
\begin{enumerate}
    \item \textbf{Cross-TX Minimization}: Reduce the number of cross-shard transactions by grouping related transactions into a same shard.
    \item \textbf{Temporal Balancing}: To distribute load evenly among shards to increase parallelism and reduce queuing time. 
\end{enumerate}


In practice, we aim to deploy OptChain as a user-side software. By monitoring its own transactions as well as the information on the loads and confirmation time at the shards, a client software can make decision on which is the best shard to submit transactions. It is important that users do not have to store the whole blockchain to optimize the placement.

In the next section, we will propose a lightweight, yet, efficient solution, answering the question: \textit{How to identify an appropriate shard for a new transaction such that in the long future, our system can achieve four main goals as described?}

\section{OptChain Algorithm} \label{sec:optchain}

In this section, we propose an algorithm used by OptChain to place transactions into shards. First, we introduce the system model, in which we represent transactions under a directed network. Under this model, we utilize a well-known PageRank analysis to propose T2S-score, which is to measure how likely a transaction should be placed into the shard. Then, we propose a mathematical model to estimate confirmation latency for placing a transaction into shards, which we called L2S score. Finally, we describe how OptChain places transactions into shards based on the combination of T2S and L2S scores.


\subsection{Transaction-as-Node Network and Partitioning}


To observe the relation between transactions, we model the set of transactions under a graph representation, which is defined as follows:


\begin{definition}[TaN Network]
A TaN network of a set of transactions is presented as a directed graph $G = (V,E)$ where $V$ is the set of transactions and $E$ is a multiset of directed edges in which the number of appearances of an edge $(u, v) \in E$ indicates the number of UTXO(s) that the transaction $v$ uses from output of transaction $u$.
\end{definition}

To have a clear picture on how a TaN network looks like, we construct a TaN network from a set of transactions in the first 705,066 blocks of the Bitcoin blockchain which were obtained by using Bitcoin Core \cite{bitcoincore}. The most recent transaction of the dataset was issued in October 2021. From this dataset, we construct a TaN network of 678,339,587 nodes and 1,726,949,132 edges. Nodes that do not have any outgoing edges represent coinbase transactions (rewards for mining Bitcoin blocks). On the other hand, nodes that do not have any incoming edges represent transactions whose UTXOs have not been spent. TaN network is a directed acyclic graph since a transaction only uses UTXO(s) of past transactions. Therefore, TaN network can be sorted in a topological order, which exactly reflects the order of appearance of transactions.




\begin{figure*}
	\centering
	\subfloat[Degree distribution]{
	    \includegraphics[width=0.31\linewidth]{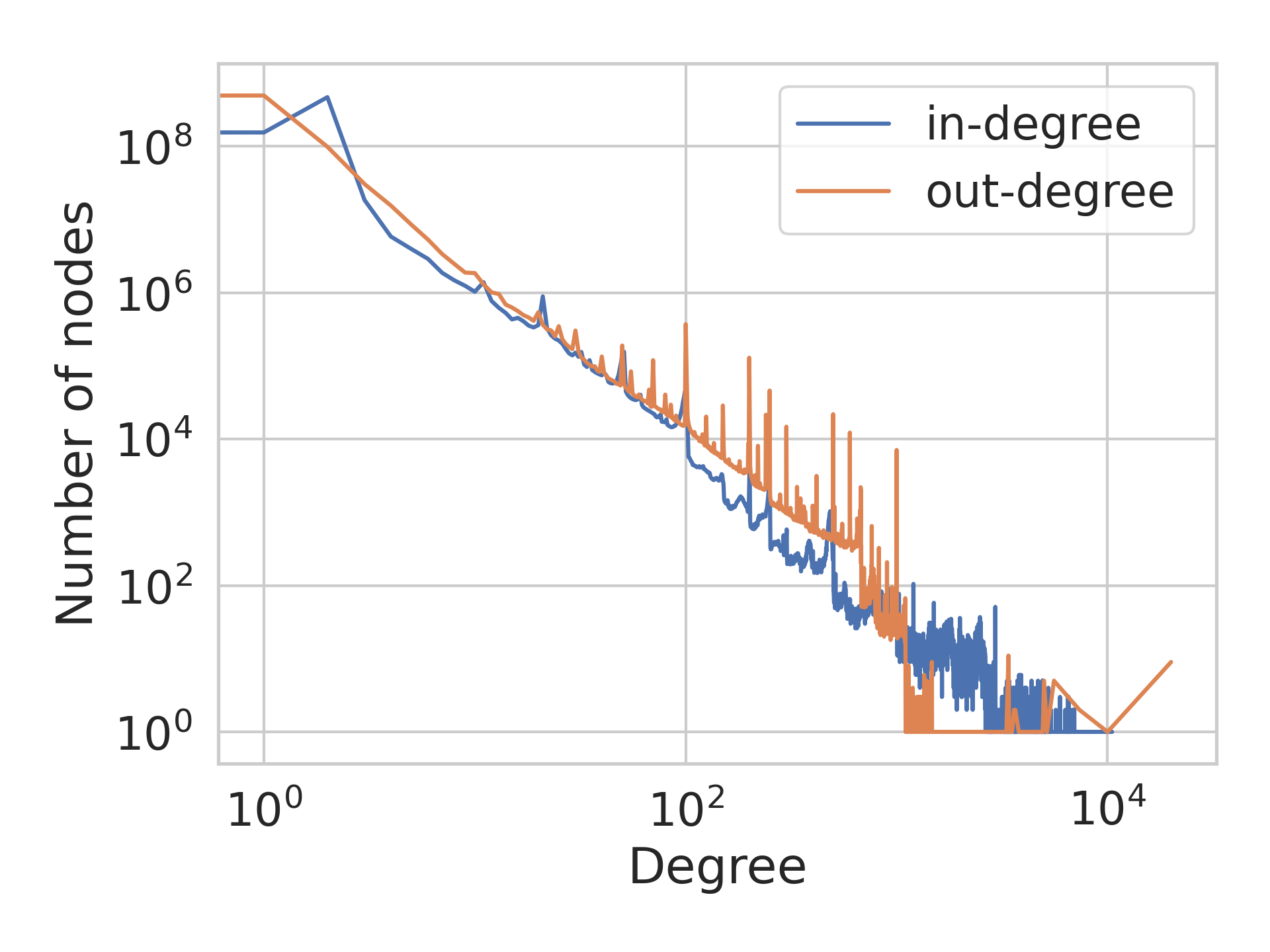}
	    \label{fig:degree-dist}
	}	
	\hfill
	\subfloat[Cumulative distribution]{
	    \includegraphics[width=0.31\linewidth]{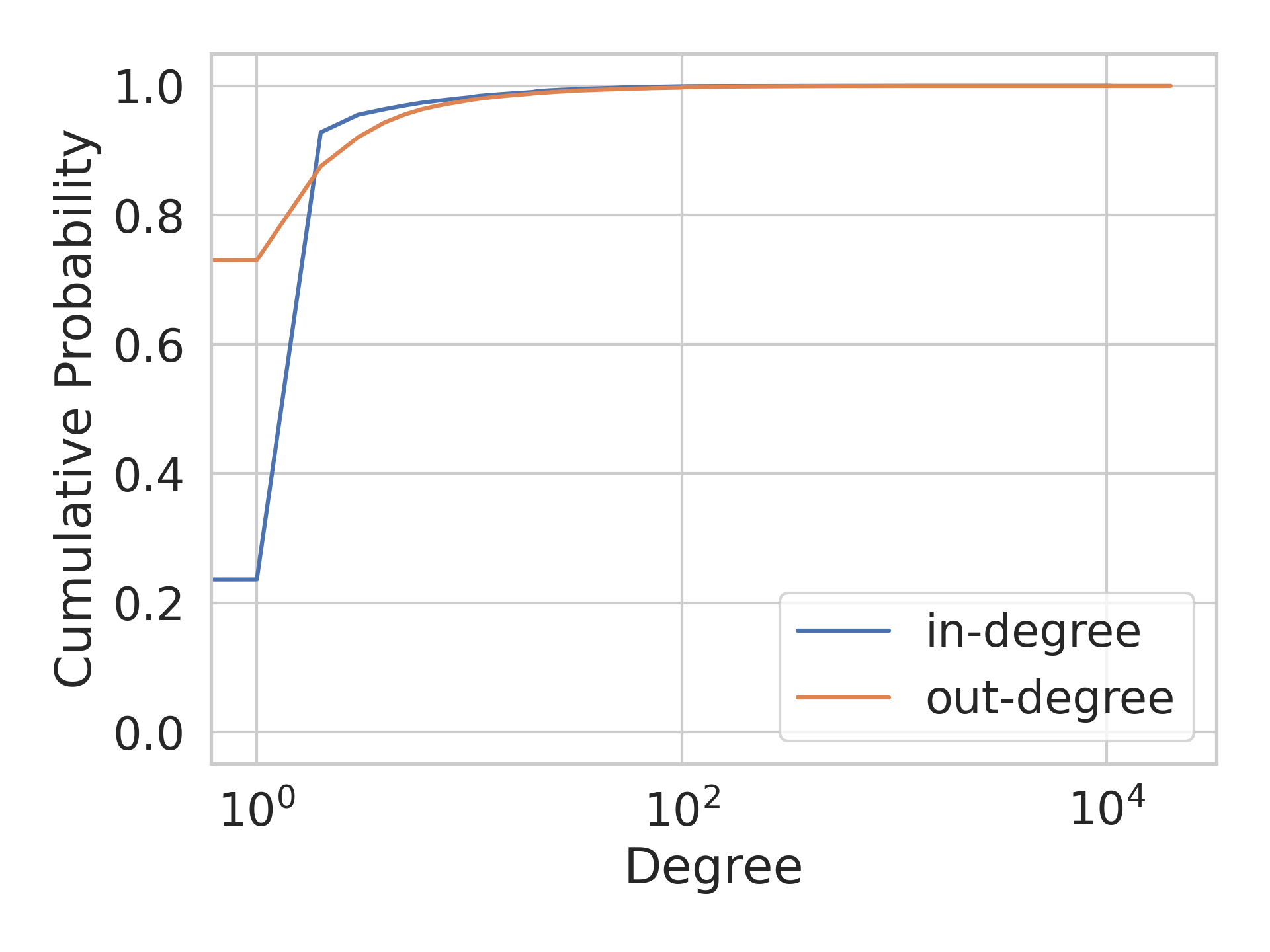}
    	\label{fig:degree-cdf}
	}
	\hfill
	\subfloat[Changes in average degree]{
	    \includegraphics[width=0.31\linewidth]{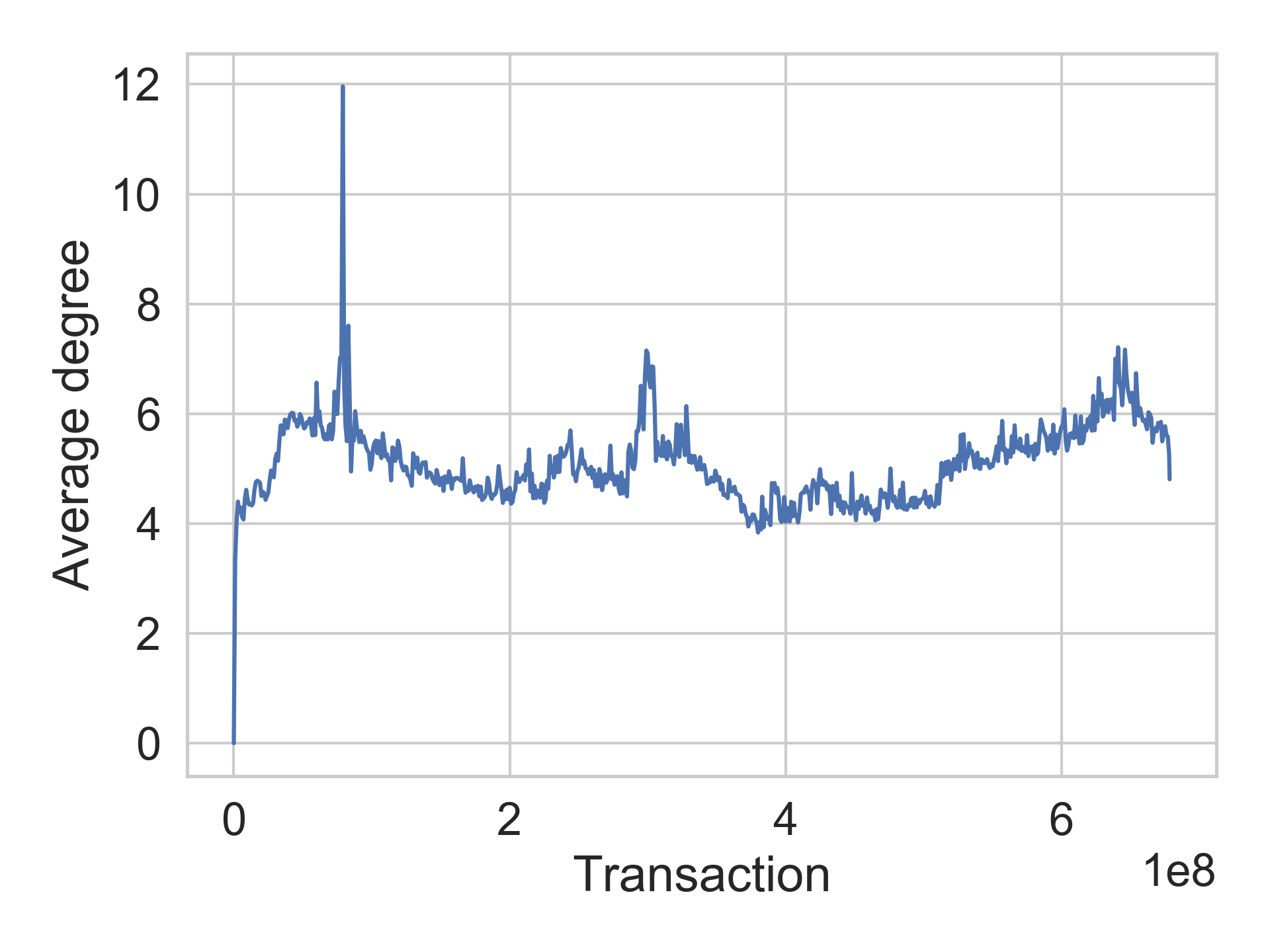}
	    \label{fig:degree-time}
	}
	\caption{TaN network statistics}
	\label{fig:stat}
\end{figure*}
Fig. \ref{fig:degree-dist} shows the plot of the degree distribution of the constructed TaN network in log-log scale. The degree distribution of TaN network exhibits \textit{power-law} distribution with average in- or out- degree of $\approx 2.5$.  There are a total of 705,066 coinbase transactions (transactions without inputs), and 5,436,269 transactions whose UTXO(s) have not been spent. Also, there are 39,787 transactions without any outputs or inputs. As can be seen from the degree distribution in \cref{fig:degree-cdf}, the vast majority of nodes (92.8\%) in the TaN network have in-degree lower than 3. As regards the out-degree, 97.5\% of nodes have the out-degree lower than 10, and 87.6\% lower than 3.


 We investigate the average degree of nodes of TaN network overtime and plot in \cref{fig:degree-time}. We observe that for most of the time, the average degree of TaN network is stable and consistent. There are only 2 periods during which the average degree significantly changes, which are the first 1 millions transaction and around the 80,000,000th one. This behavior does not happen regularly because: (1) the first period is that the system needs to generate a lot of coinbase transactions for funding and (2) there was a flooding attack at the second period \cite{pearson_2015}, which causes mining pools to create a lot of transactions with high degree to clean up ``trash" transactions.

By distributing transactions into shards, the task of transactions sharding eventually become partitioning the TaN network into $k$ disjoint subsets of nodes $\mathcal{S} = \{S_1,...,S_k\}$, where $k$ is the number of shards and $S_i$ denotes a set of transactions under management of shard $i$. We have $\cup_{i=1}^k S_i = V$ and $S_i \cap S_j = \emptyset$ for all $i,j = 1\rightarrow k$. For simplicity, we also call $S_i$ as the shard $i$.

Given a transaction represented by node $u \in V$, denote $S(u) \in \mathcal{S}$ as a shard containing $u$. Let $\mathcal{S}_{in}(u)$  as a set of input shards of $u$, i.e $\mathcal{S}_{in}(u) = \{ S_i \mid \exists v \in S_i, (v,u) \in E \}$. Then $u$ is a \crsh{} iff $\mathcal{S}_{in}(u) \neq \{S(u)\}$. Therefore, we consider the task of distribute a new transaction into shards as an online partitioning to the TaN network. To be specific, given a TaN network $G=(V,E)$, a set of $k$ disjoint subsets of nodes $\mathcal{S} = \{S_1,S_2,...,S_k\}$ and a new arriving transaction (node) $u$, our task is to identify $S(u) \in \mathcal{S}$.  



\subsection{Transaction-to-Shard Score} \label{subsec:t2s}

In this part, we will propose a metric to measure the fitness score between the new arrival transaction and shards, which we called Transaction-to-Shard (T2S) score. This score is motivated by a well-known graph measurement, called PageRank. To be specific, PageRank assigns numerical weighting to each node of a network with the purpose of measuring node relative importance within the network. A PageRank vector is a weighted sum of the probability distribution obtained by taking a sequence of random walk steps starting from a specified initial distribution. PageRank has been proven to be an efficient method on graph local partitioning, in which nodes in a same set of partition tend to have a similar weight.

Although regular PageRank computation could cost considerable runtime, we utilize the trait that TaN is acyclic and the order of nodes' appearance is also TaN's topological order to devise a fast T2S-score computation  of each nodes. To be specific, we represent the T2S-scores of a transaction $u$ to shards under a $k$-dimensional vector $p(u)$, in which the entry $i$ measures the fitness between $u$ and shard $i$. Denote $r[i]$ as the value of entry $i$ of vector $r$. Also, let $N_{in}(u)$ be a multiset of input transactions of $u$, where the number of appearances of a transaction corresponds to the number of UTXO(s) that $u$ spent from it. On the other hand, $N_{out}(u)$ is a multiset of output transactions of $u$. $p(u)$ is computed by the following equation:

\begin{align*}
    p(u) = \alpha s(u) + (1-\alpha) \sum_{v \in N_{in}(u)} \frac{p(v)}{|N_{out}(v)|}
\end{align*}
where $\alpha$ is a constant in $(0,1]$ and $s(u)$ is a starting vector for a node $u$. $s(u)$ is initiated as follows:

\begin{itemize}
    \item If $u$ is a new arriving node, $s(u) = \{0\}^k$
    \item Otherwise, $s(u)$ receives a value $\frac{1}{|S(u)|}$ at the entry corresponding to shard $S(u)$ and $0$ elsewhere.
\end{itemize}

Clearly $p(u)$ can easily be calculated in the manner similar to Bread-First Search, in which we start from nodes who have no input (coinbase transaction). Therefore, computing $p(u)$ for all $u \in V$ costs $O(k(|V| + |E|))$ runtime complexity, which is very expensive as the TaN network grows. 
However, we observe that: after placing a new transaction $u$ into shard $S(u)$, the change on $s(v)$ for all $v \in V$ does not impact the fitness score of all nodes other than normalization scale. To be specific, assume at time $t+1$,  $u$ arrives and is placed into shard $i$ ($S(u) = S_i$). Denote $p^{(t)}(v)$ is the T2S-score vector of node $v$ at time $t$ and $S^{(t)}_i$ is the shard $i$ at time $t$. We have:
\begin{itemize}
    \item If $v \neq u$, $p^{(t+1)}(v)[i] = \frac{|S^{(t)}_i|}{|S^{(t)}_i| + 1} p^{(t)}(v)[i]$ and $p^{(t+1)}(v)[j] =  p^{(t)}(v)[j]$ for all $j \neq i$.
    \item If $v = u$, $p^{(t+1)}(v)[i] = \alpha \frac{1}{|S^{(t)}_i| + 1} +  \frac{S^{(t)}_i}{S^{(t)}_i + 1} p^{(t)}(v)[i]$ and $p^{(t+1)}(v)[j] =  p^{(t)}(v)[j]$ for all $j \neq i$.
\end{itemize}

Therefore, rather than computing $p(v)$ for all $v \in V$ from scratch to get T2S score of a new transaction $u$, we propose the following methods for the faster calculation. First, we introduce two other vectors $p^\prime(v), s^\prime(v)$ associated to each node $v \in V$ in addition to $s(v)$ and $p(v)$. If a transaction $v$ is placed in shard $j$ then $s^\prime(v)[j] = 1$ and $s^\prime(v)[i] = 0$ for all $i \neq j$, $s^\prime(v)$ is fixed right after $v$ is placed. Given a new transaction $u$, $p(u)$ can be computed as follows.

\begin{itemize}
     \item If $u$ is a coinbase transaction, there is no calculation needed. 
    \item Otherwise, we set $p^\prime(u) = (1-\alpha) \sum_{v \in N_{in}(u)} \frac{p^\prime(v)}{|N_{out}(v)|}$. The T2S score of node $u$ is $p(u) = \{\frac{p^\prime(u)[i]}{\sum_v s^\prime(v)[i]}\}^k_i$. 
\end{itemize}
Then, after placing $u$ into shard $i$ ($S(u) = S_i$), we update $p^\prime(u) = p^\prime(u) + \alpha s^\prime(u)$. Overall, the computation of $p(u)$ now only costs $O(|N_{in}(u)|k)$. As TaN network has been shown to be scale-free, the average computation, thus, costs only $O(k)$.

\textbf{Discussion.} Why do we develop the new T2S score instead of using existing graph partitioning methods to minimize the number of \crsh{}? The best way yet unrealistic to minimize the number of \crsh s is that we know the structure of the TaN network beforehand, i.e. all transactions have been arrived, and apply graph partitioning algorithms. But even those, does such algorithm improve the performance of sharding blockchain? The answer could be not likely. In experiment, we applied a well-known graph partitioning tool, called \textbf{Metis} $k$-way \cite{karypis1995metis}, to get the partitioning on transactions. Metis aims at partitioning the graph into disjoint sets of nodes with almost equal size while minimizing the number of edges whose two endpoints belong to different sets. However, when simulating, if we put transaction exactly like in the Metis solution, the system's throughput and confirmation latency are highly impacted as the Metis solution tends to put large amount of consecutive transactions into one shard. More details of such experiments will be shown in Section \ref{sec:experiment}.

\DeclarePairedDelimiter\floor{\lfloor}{\rfloor}
A more realistic method, but simple, is \textbf{Greedy}. To compare with Metis, given $n$ txs and $k$ shards, we set the maximum number of transactions that a shard can have as $(1+\epsilon)\floor{n/k}$. Then we consider transactions sequentially. Considering a transaction $u$ arrives at a moment $t$, denote $S^{(t)}_i$ as a set of transactions in shard $i$ at time $t$. The greedy algorithm computes the cost of placing $u$ on shard $j$ as $f(u,j) = |\mathcal{S}_{in}(u) \setminus S^{(t)}_j|$. Then the algorithm places $u$ into a shard $j$ which has the maximum $f(u,j)$ and its size has not exceeded $(1+\epsilon)\floor{n/k}$. Intuitively, the greedy solution will help reduce the number of \crsh s. However, this solution does not take the global view on TaN network structure while only considering connection of one-hop away from $u$. Thus, in the long future, the performance of greedy algorithm becomes undesireable.

To prove the efficiency of T2S-score, we compare the percentage of \crsh s between Omniledger, Greedy solution and a solution which is similar to Greedy except that after computing $p(u)$, we put $u$ into a shard with the highest T2S-score, i.e. $\mbox{argmax}_i p(u)[i]$. We set $\alpha = 0.5$ and we call such method \textbf{T2S-based}. With both Greedy and T2S solution, we set $\epsilon = 0.1$.

We use the same data set from MIT. First, we ran the algorithms from scratch where all shards are empty at the beginning. The result is presented in Table \ref{table:crosstx}. It is easy to see that our solution significantly reduces the number of \crsh s and overcomes Greedy solution with a huge margin. Even Metis is unrealistic, we put its results in the table as a baseline for comparison.


\begin{table}[h]
\centering
\caption{Percentage of \crsh s when running from scratch} \label{table:crosstx}
\begin{tabular}{c | c | c | c | c }
\toprule
$k$ & Metis & Greedy & Omniledger & T2S-based \\ 
 \midrule
4 & 1.66 \% 	&  24.62 \% & 80.82 \% & 9.28 \%\\
8 & 3.09 \% 	&  27.02 \% & 90.33 \% & 12.52 \%\\
16 & 4.70 \% 	&  28.14 \% & 94.87 \% & 15.73 \%\\
32 & 6.91 \% 	&  28.69 \% & 97.09 \% & 18.94 \%\\
64 & 9.91 \% 	&  28.97 \% & 98.18 \% & 21.65 \%\\
\bottomrule
\end{tabular}
\end{table}

Next, we consider at a certain moment, the system already places a certain amount of transactions into shards. We then apply the algorithms with a set of new arrival transactions and compare the number of \crsh s in such set. To be specific, first, we use Metis to partition the TaN network of 30 millions Bitcoin transactions into $k$ shards. Then we apply the algorithm to place transactions of a sequence of next 1 millions transactions into $k$ shards. The results is presented in Table \ref{table:seq_crosstx}. Again, our solution showed its efficiency comparing with Greedy and OmniLedger in term of minimizing the number of \crsh s.

\begin{table}[t]
\centering
\caption{Number of \crsh s when running from a certain stage of the system} \label{table:seq_crosstx}
\begin{tabular}{c | c | c | c }
\toprule
$k$ &  Greedy & Omniledger & T2S-based \\ 
 \midrule
4 &  335,269& 837,356 & 112,657\\
8 &   407,747& 922,073 & 172,978\\
16 &  441,267& 960,935 & 226,171\\
32 &  449,032& 979,323 & 282,108\\
64 &  454,321& 988,144 & 366,854\\
\bottomrule
\end{tabular}
\end{table}

\subsection{Latency-to-Shard Score and Temporal Fitness}

We have shown that a simple solution using T2S-score could significantly reduce the number of \crsh s. However, this is not a ultimate goal of our algorithm design. What if there is a huge amount of transactions with the same ``fittest'' shard w.r.t T2S-score arriving sequentially?  Trivially, putting all those transactions into the same shard is not a good solution. Therefore, in this part, we propose a mathematical model to estimate confirmation latency of a transaction under transaction sharding. We call the estimated latency \textit{Latency-to-Shard} (L2S) score. Hence, the best transaction placement is the one that maximizes the T2S-score while minimizing the L2S-score.


Let consider a moment a new transaction $u$ arrives and system shards are $S_1,...S_k$. Assume if $u$ is placed in shard $j$, $u$ will need proof-of-acceptance from a set of shards $\mathcal{S}_j$. We model the communication time between a user who creates $u$ and the shard $S_i$ under exponential distribution $l_c^{(i)}(t) = \lambda_c^{(i)} e^{-\lambda^{(i)}_v t}$, where $\frac{1}{\lambda_c^{(i)}}$ is expected communication time, which could be collected through frequently sampling between the user and shard $S_i$. Also, for each shard $S_i$, we model the verification time of $S_i$ under exponential distribution $l^{(i)}_v(t) = \lambda^{(i)}_v e^{-\lambda^{(i)}_v t}$, where $\frac{1}{\lambda_v^{(i)}}$ is expected verification time, which could be estimated from observation of recent consensus time of shard $i$ and its current queue size. With high precision, it is likely that $\lambda_v^{(1)} \neq ... \neq \lambda_v^{(k)} \neq \lambda_c^{(1)} \neq ... \neq \lambda_c^{(k)}$

Therefore, a probability distributed function of time to get proof-of-acceptance from shard $S_i$ is modeled as follows.

\begin{align*}
    f^{(i)}(t) & = \int_{0}^{t} l^{(i)}_c(x) l^{(i)}_v (t-x) \Delta x \\
    & = \frac{\lambda^{(i)}_c \lambda^{(i)}_v}{\lambda_v^{(i)} - \lambda_c^{(i)}} \Big( e^{-\lambda_c^{(i)} t} - e^{-\lambda_v^{(i)}t} \Big)
\end{align*}
while the cumulative distributed function is:
\begin{align*}
    &F^{(i)}(t < T) = \int_{0}^{T} \frac{\lambda^{(i)}_c \lambda^{(i)}_v}{\lambda_v^{(i)} - \lambda_c^{(i)}} \Big( e^{-\lambda_c^{(i)} t} - e^{-\lambda_v^{(i)}t} \Big) \Delta t \\
    &\quad = \frac{\lambda_v^{(i)}}{\lambda_v^{(i)} - \lambda_c^{(i)}} (1-e^{-\lambda_c^{(i)}T}) - \frac{\lambda_c^{(i)}}{\lambda_v^{(i)} - \lambda_c^{(i)}} (1-e^{-\lambda_v^{(i)}T}) 
\end{align*}

\begin{algorithm}[t]
	\caption{Transaction Sharding in OptChain}\label{alg:onlinealg}
	\textbf{Input}: $G(V,E)$, $S_1,...S_k$, $p^\prime: V \rightarrow \mathbb{R}^+$, $\alpha$ and a new transaction $u$ \\
	\textbf{Output}: a new state of $S_1,...S_k$
	\begin{algorithmic}[1]
		    \State \textcolor{blue}{\# Compute T2S-scores of $u$}
		    \State $p^\prime(u) = (1-\alpha) \sum_{v \in N_{in}(u)} \frac{p^\prime(v)}{|N_{out}(v)|}$
		    \State $p(u) = \{\frac{p^\prime(u)[i]}{|S_i|}\}_{i=1}^k$
		    \State \textcolor{blue}{\# Compute L2S-scores of $u$}
		    \For{$j = 1 \rightarrow k$}
		        \State $\mathcal{E}(j) = \int_{0}^\infty t \int_0^t f_v^{(j)}(x) f_v^{(j)}(t-x) \Delta x \Delta t$
		    \EndFor
		    \State \textcolor{blue}{\# Put $u$ into shard with the highest Temporal Fitness score}
		    \State $s_u = \textnormal{argmax}_{i} \Big(p(u)[i] - 0.01 \cdot \mathcal{E}(i) \Big)$
		    \State $S_{s_u} \leftarrow S_{s_u} \cup \{u\}$
		    \State \textcolor{blue}{\# \textit{Update after placing $u$ into $S_{s_u}$}}
		    \State $p^\prime(u)[s_u] = p^\prime(u)[s_u] + \alpha$
		\State \textbf{Return } $S_1, ...S_k$
	\end{algorithmic}
\end{algorithm}

As the user can send request for verification simutaneously to shards in $\mathcal{S}_j$, the probability that verification process is done by time $T$ is computed as
\begin{align*}
    &F(t < T) = \prod_{S_i \in \mathcal{S}_j} F^{(i)}(t < T) 
\end{align*}

Thus, the probability distributed function of time for the user to get all proof-of-acceptance if place $u$ into shard $j$ is modeled as follows
\begin{align*}
    &f_v^{(j)}(t) = \frac{\Delta F(t)}{\Delta t} \\
    &= \sum_{S_i \in \mathcal{S}_j} \frac{\lambda^{(i)}_c \lambda^{(i)}_v}{\lambda_v^{(i)} - \lambda_c^{(i)}} \Big( e^{-\lambda_c^{(i)} t} - e^{-\lambda_v^{(i)}t} \Big) \prod_{S_r \in \mathcal{S}_j \setminus S_i} F^{(r)}(t)
\end{align*}

Similarly, we can find the probability distribution for $u$ to get confirmation from shard $j$, which is:
\begin{align*}
    f_c^{(j)}(t) = \frac{\lambda^{(j)}_c \lambda^{(j)}_v}{\lambda_v^{(j)} - \lambda_c^{(j)}} \Big( e^{-\lambda_c^{(j)} t} - e^{-\lambda_v^{(j)}t} \Big)
\end{align*}

Therefore, the L2S-score of $u$ if putting into shard $j$ is computed by
\begin{align*}
    \mathcal{E}(j) = \int_{0}^\infty t \int_0^t f_v^{(j)}(x) f_v^{(j)}(t-x) \Delta x \Delta t
\end{align*}

In overall, for a given transaction, we would like to balance between the T2S and L2S-score. Specifically, given a new arrived transaction $u$ and a shard $j$, we define the \textit{Temporal Fitness} score between $u$ and $j$ as $p(u)[j] - 0.01 \cdot \mathcal{E}(j)$. OptChain then places $u$ into the shard whose has the highest Temporal Fitness score. The procedure of transaction sharding in OptChain is presented in detailed by Alg.  \ref{alg:onlinealg}.

\section{Experiments} \label{sec:experiment}

The aim of this experiment is to evaluate the impact of OptChain on Blockhain system's latency and throughput. To demonstrate the strengths of our proposed solution, the performance of OptChain is rigorously compared to OmniLedger, Metis, and the Greedy heuristic discussed in section \ref{sec:optchain}.

\subsection{Experiments Design and Configuration}
The experiments presented in OmniLedger \cite{kokoris2017omniledger} used the first, arguably simple, 10,000 Bitcoin blocks which contain only 10,093 transactions in total. Additionally, of those transactions, 99.1\% are coinbase, which cannot be \crsh s. At that point in time, Bitcoin was still in an initial phase in which there was only about 1 transaction per block and most of them were coinbase transactions. Therefore, their experimental results do not represent the impact of \crsh s.

Our experiments are conducted on the first 10 million Bitcoin transactions, taken from the Bitcoin Core client \cite{bitcoincore}. From this dataset, we build a TaN network that contains 10,000,000 nodes and 19,958,051 edges. As opposed to OmniLedger, our dataset is much bigger with more chance of creating \crsh s. Specifically, with 16 shards, the OmniLedger's random placement produces 9,349,979 \crsh s comprising 93.5\% of the whole set.


We develop a sharding-based blockchain simulation and run the transactions placement algorithms to calculate the latency and throughput. Particularly, we use the OverSim framework \cite{OverSim_2007} to simulate a Bitcoin-like blockchain system on OMNeT++ 4.6 \cite{varga2008overview}, which is a discrete event-based network simulator. The bandwidth of all connections between nodes is set to 20 Mbps and a latency of 100 ms is imposed on all communication links. We set the block size of 1MB because this is currently the block size limit in Bitcoin. The average size of a transaction is about 500 bytes, so we put about 2000 transactions in one block. A shard is assigned with about 400 validators and one leader that are randomly placed at different coordinates. In our simulation, the distance between nodes affects the communication latency. Each shard implements a queue (or mempool) to store incoming transactions that have not been processed yet.

Our simulation involves a set of clients that continuously issue transactions from the dataset to the system at a predefined rate. We re-implement the mechanism for handling \crsh s as decribed in Section \ref{sec:blockchain-sharding}. Before sending a transaction $u$ for verification, a client will run a transactions placement algorithm to determine the shard $S(u)$ to place that transaction.

\textbf{Getting rid of Omniledger's bottleneck.} In Omniledger, users gossip each transaction to all nodes in the network, thus, all nodes must receive the whole blockchain (and also the unconfirmed transactions, if any). Thus, even when the number of nodes and shards go to infinity, the throughput of nodes in Omniledger is still limited by typical network bandwidth of the nodes. To address this issue, users will direct each transaction directly to the input shards and output shards of the transaction.

\textbf{Transaction placement strategies.} We compare \optch{} with three transaction placement methods. 
\begin{itemize}
    \item OmniLedger: The default random placement strategy in Omniledger.
    \item Greedy: Greedy placement strategy, discussed in Section~\ref{subsec:t2s}.
    \item Metis $k$-way: Running (offline) Metis partitioning algorithm to partition the network into $k$ shards.
\end{itemize}
As Metis $k$-way is an offline algorithm, it is not a realistic transactions sharding scheme. However, if we can put transactions as in Metis solution, we can minimize the number of \crsh s. Our purpose is, therefore, to see whether only minimizing the number of \crsh s can improve the system performance. Therefore, we first input the whole TaN network to get its Metis solution and then use the resulting partitions to determine $S(u)$ for each transaction $u$.

We carry out the experiment with various configurations by combining different number of shards and transaction rates. Specifically, we vary the number of shards from 4 to 16 to conform with the experiments presented in OmniLedger \cite{kokoris2017omniledger}. With regard to the transactions rate, which represents the rate at which transactions are sent to the system, we refer to the VISA-level throughput of 4000 transactions per second (tps) to set the transactions rate from 2000 tps to 6000 tps. \cref{tab:ex} summarizes all the parameters and configuration of the experiments.

\begin{table}[]
\caption{Experiment configuration}
\label{tab:ex}
\begin{tabular}{@{}ll@{}}
\toprule
Number of transactions  & 10,000,000                                \\
Block size              & 1 MB                                      \\
Transactions per block  & 2,000                                     \\
Network bandwidth       & 20 Mbps                                   \\
Number of shards        & 4, 6, 8, 10, 12, 14, 16                       \\
Transactions rate (tps) & 2000, 3000, 4000, 5000, 6000  \\
Algorithms              & OptChain, Metis k-way, OmniLedger, Greedy \\
 \bottomrule
\end{tabular}
\end{table}


\begin{figure}[t]
    \centering
    \subfloat[OptChain]{
        \includegraphics[width=.5\linewidth]{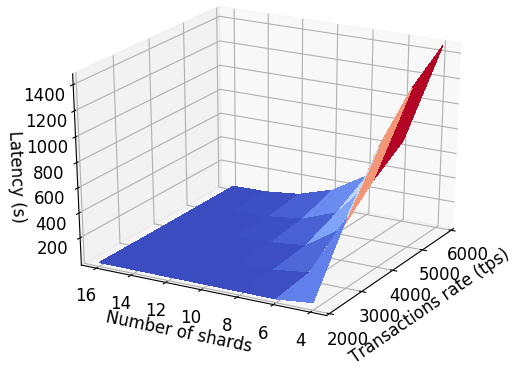}\hfill
        \includegraphics[width=.5\linewidth]{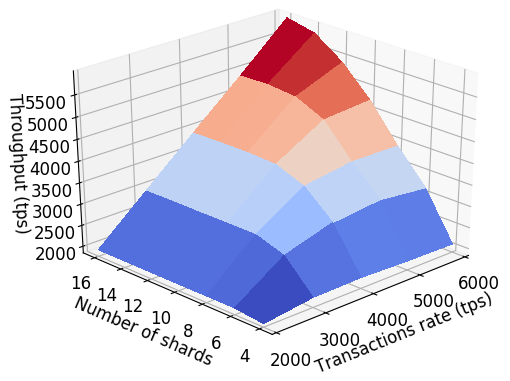}
        \label{fig:optchain}
    }\\
    \subfloat[OmniLedger]{
        \includegraphics[width=.5\linewidth]{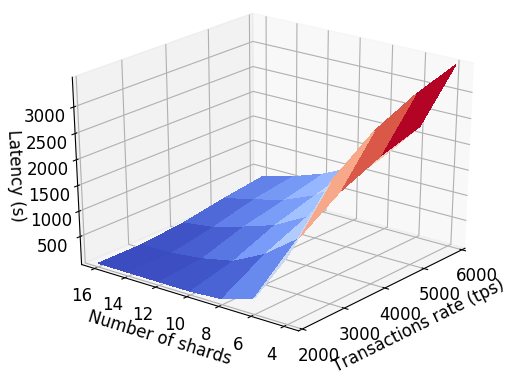}\hfill
        \includegraphics[width=.5\linewidth]{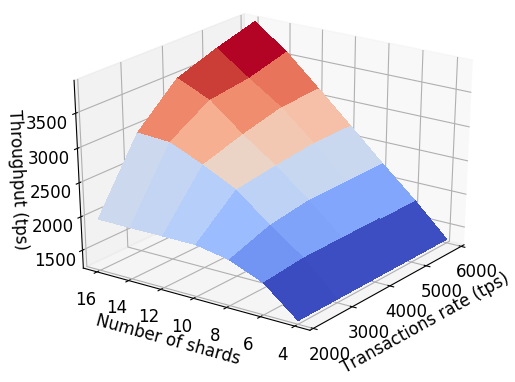}
    }\\
    \subfloat[METIS k-way]{
        \includegraphics[width=.5\linewidth]{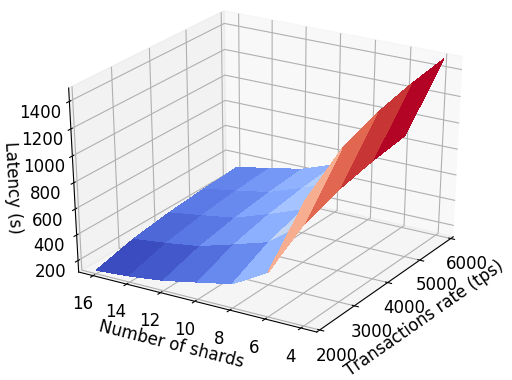}\hfill
        \includegraphics[width=.5\linewidth]{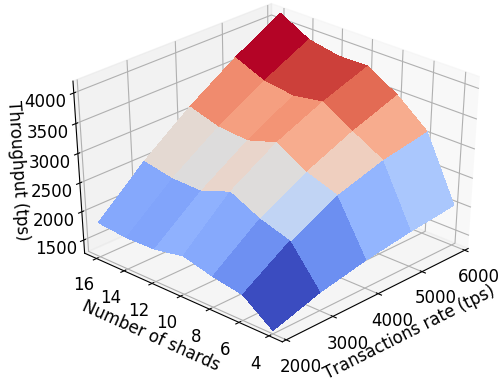}
    }\\
    \subfloat[Greedy]{
        \includegraphics[width=.5\linewidth]{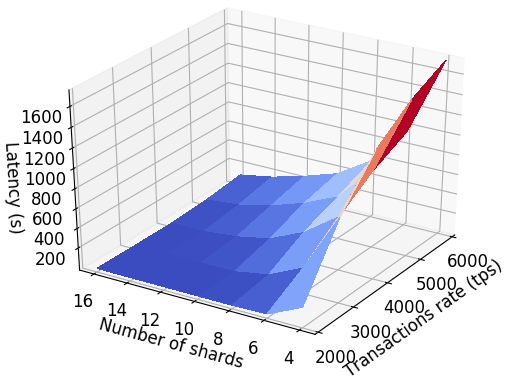}\hfill
        \includegraphics[width=.5\linewidth]{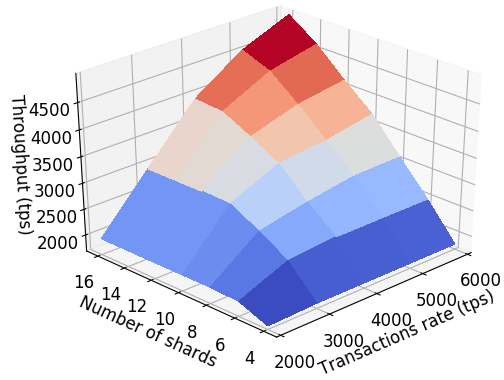}
    }
    \caption{Impact of different transactions rates and number of shards on the latency and throughput}
    \label{fig:impact}
\end{figure}

\subsection{Experimental results}


In \Cref{fig:impact}, we summarize the system's average latency and throughput when running the algorithms with different combination of transactions rates and number of shards. The thoughput is calculated by taking the number of transaction divided by the total time for all transactions get committed. The latency of a transaction is measured by the time from when the transaction is sent until it is committed to the blockchain. In this part, we would like to have more insight on how different configuration impacts the performance of OptChain as well as other transaction sharding methods in term of throughput and latency.

\subsubsection{Maximum Throughput}

From \cref{fig:impact}, it is easy to see that all the methods achieve their highest throughput at transaction rate of 6000 and 16 shards. However, except OptChain, the other three algorithms produce throughput much lower than the input rate, which shows that these three are incapable of handling such transaction rate in this setting. We observe that even OptChain does not always run well in all configurations, i.e. throughput is lower than transaction rate. But for each value of transaction rates, OptChain always has a certain configuration on the number of shards guaranteeing no backlogging in the system. To be specific, with a transaction rate of 2000, OptChain is totally healthy with at least 6 shards. This number in transaction rates of 3000, 4000, 5000, 6000 is 8, 10, 14, 16 respectively. Meanwhile, the Omniledger system needs at least 16 shards to be able to process up to 3000 transactions per second. With 16 shards, Greedy is only able to process with transactions rate up to 5000. Therefore, given a same transaction rate, OptChain needs a lower number of shards than other methods to avoid backlogging in the system.

\begin{figure}
	\centering
	\hspace{-20px}
	\subfloat[Number of shards = 16]{
	    \includegraphics[width=0.5\linewidth]{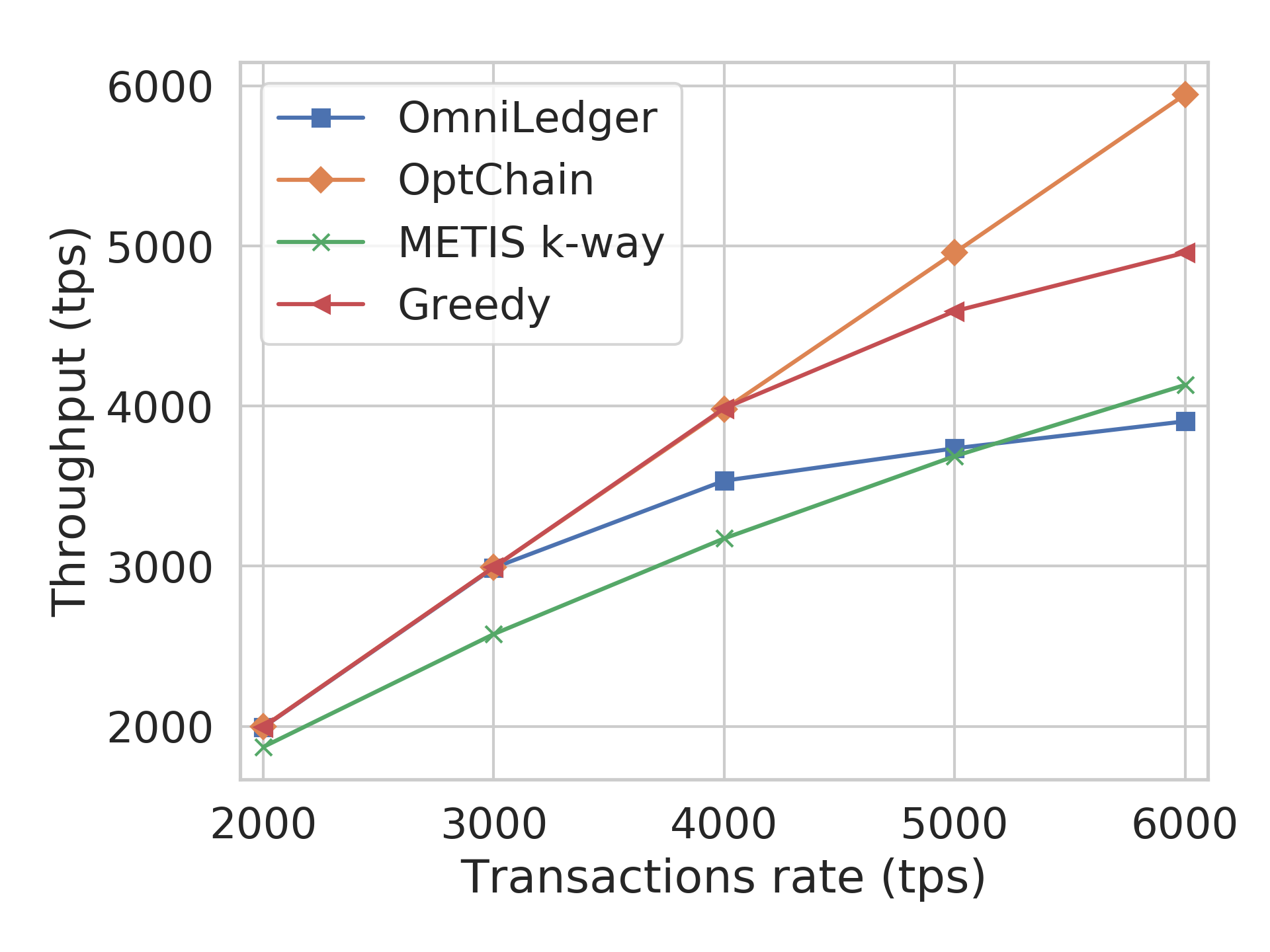}
    	\label{fig:thru-rate}
	}
	\subfloat[Varying transactions rate and \#shards]{
	    \includegraphics[width=0.5\linewidth]{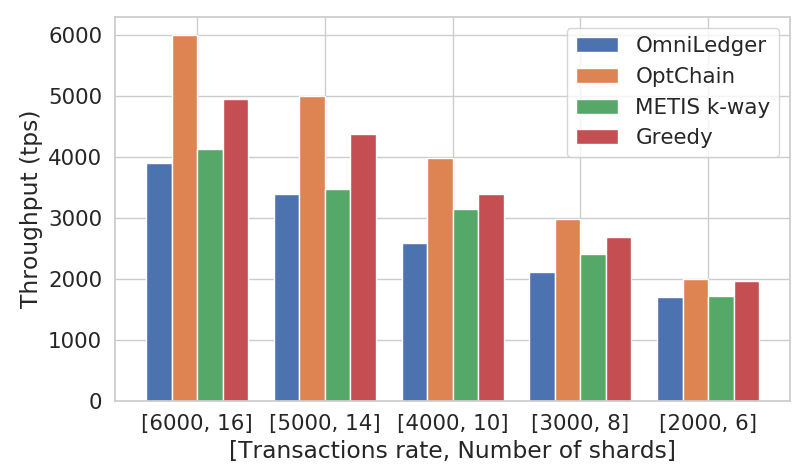}
	    \label{fig:thru-shard}
    }
	\caption{System throughput}
	\label{fig:thru}
\end{figure}

For a more comprehensive comparison, \cref{fig:thru-shard} presents the maximum system throughput at different pairs of value of transactions rate and number of shards. As can be seen, no other transaction sharding methods can reach up to the same level as OptChain. For example, the maximum throughput OptChain can achieve at 16 shards is 34.4\%, 30.5\%, and 16.6\% higher than that of OmniLedger, Metis, and Greedy, respectively.



Metis has been proven to be the best method, yet unrealistic, to distribute transactions into shards in order to minimize the number of \crsh s. High number of \crsh s is claimed by Omniledger \cite{kokoris2017omniledger} to be the main factor limiting the system performance. However, when we place transactions as in Metis's solution, the throughput never inline with transaction rate. For example, Fig. \ref{fig:thru-rate} shows the throughput of the algorithms as we fix the number of shards to 16 and vary the transactions rates. As previously stated, OptChain, Omniledger and Greedy are comfortable to a certain rate within such range. But Metis's throughput never reaches the transaction rate. Thus, we conclude that \textit{high number of \crsh s is not a sole factor hindering Blockchain sharding performance.} 

\begin{figure}
	\centering
	\includegraphics[width=0.7\linewidth]{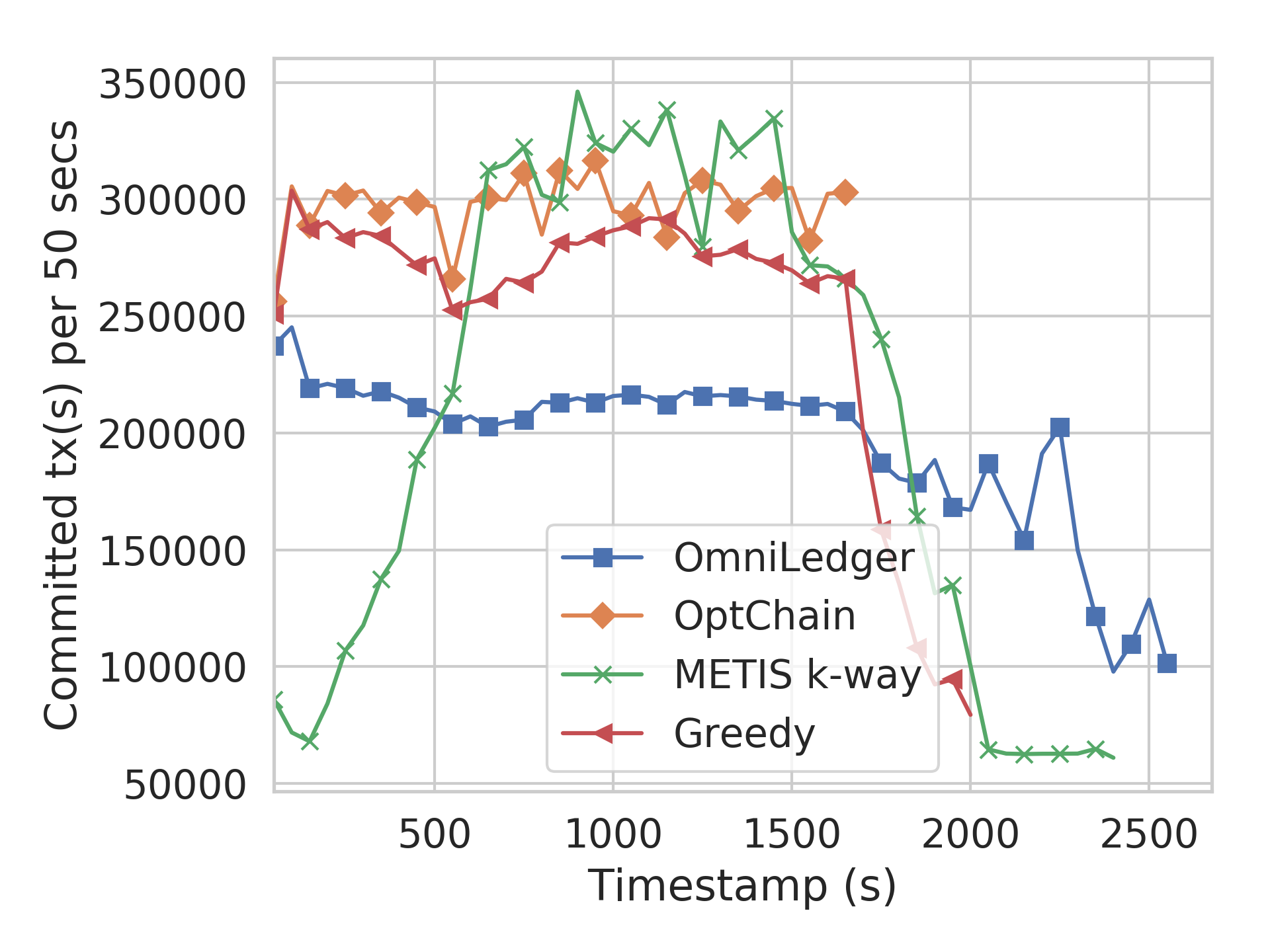}
	\caption{Number of committed transactions across time}
	\label{fig:thru-time}
\end{figure}

What cause Metis's throughput being worse than other three solutions? We take more insights on the timeline that transactions are finally committed. We set the transaction rate to be 6000 and 16 shards. Then we count the number of transactions getting committed in each 50-seconds period and plot the results as in Fig. \ref{fig:thru-time}. It is easy to see that OptChain, OmniLedger, and Greedy produce almost consistent number of committed transactions over each period of 50 seconds. Meanwhile Metis is not efficient during the first 500 seconds. Also, Metis tends to agitate more than other three solutions, which could imply congestion on shards in some certain moments, i.e. some shards get more transactions than the other. The huge drop in the end of each line in Fig. \ref{fig:thru-time} is because the simulator has reached the end of the dataset in which no more transactions are sent to the system. 

\begin{figure}[t]
\subfloat[OptChain]{\includegraphics[width=0.25\textwidth]{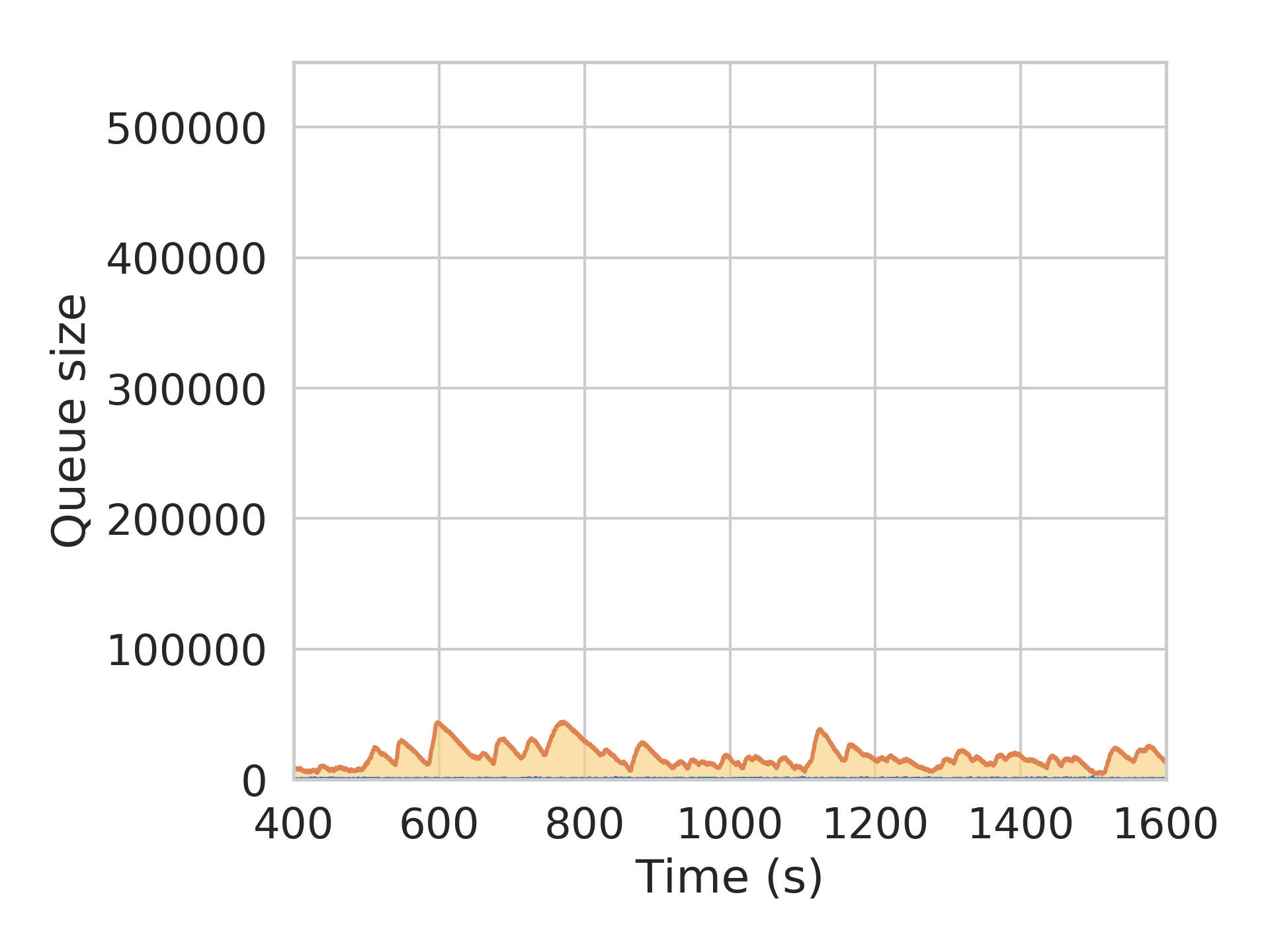}
\label{fig:queue_pr}}
~
\hspace{-20px}
 \subfloat[Omniledger]{
  	\includegraphics[width=0.25\textwidth]{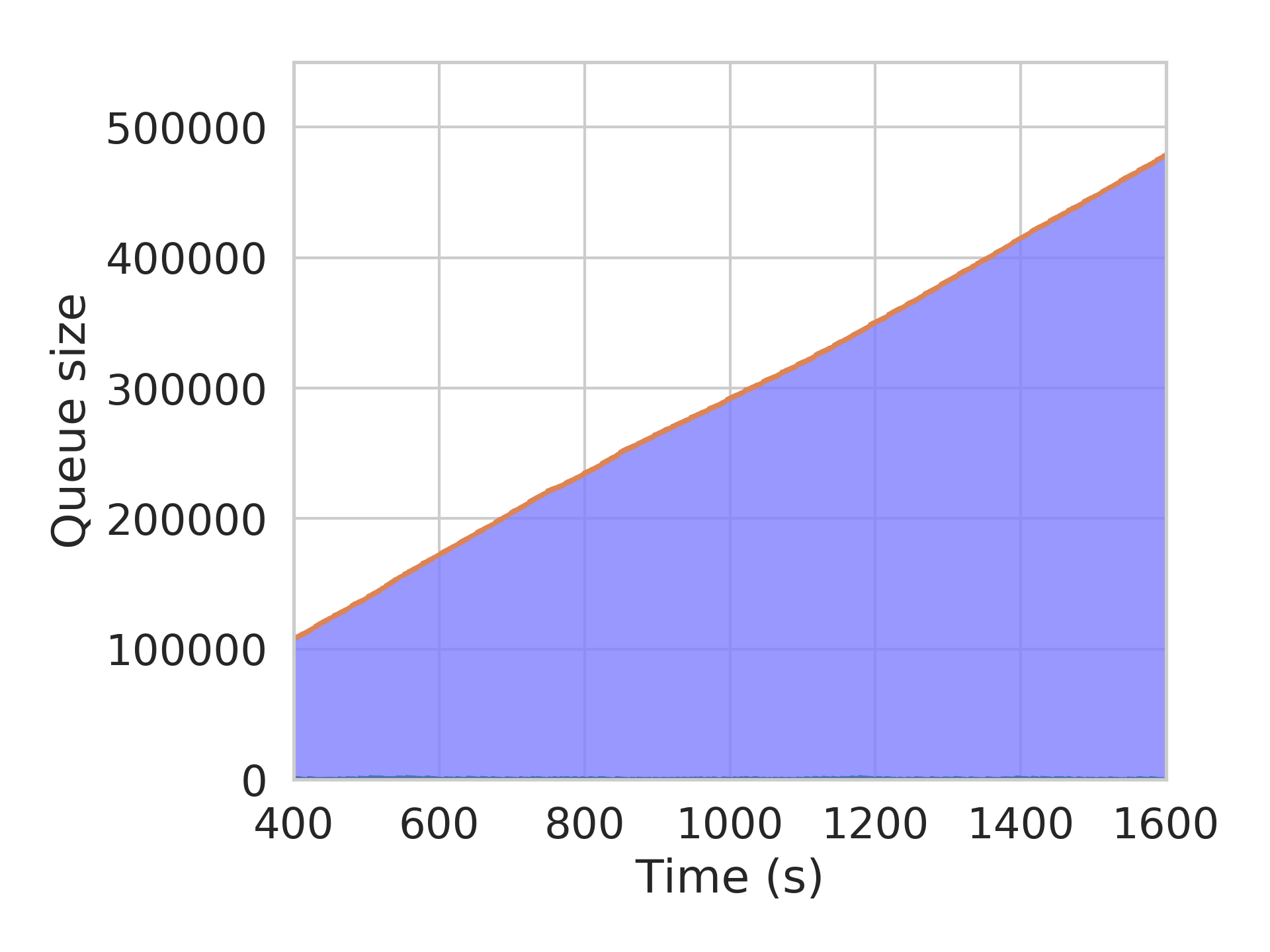}
  	\label{fig:omni_queue}}
~
\\
\subfloat[Greedy]{
 	\includegraphics[width=0.25\textwidth]{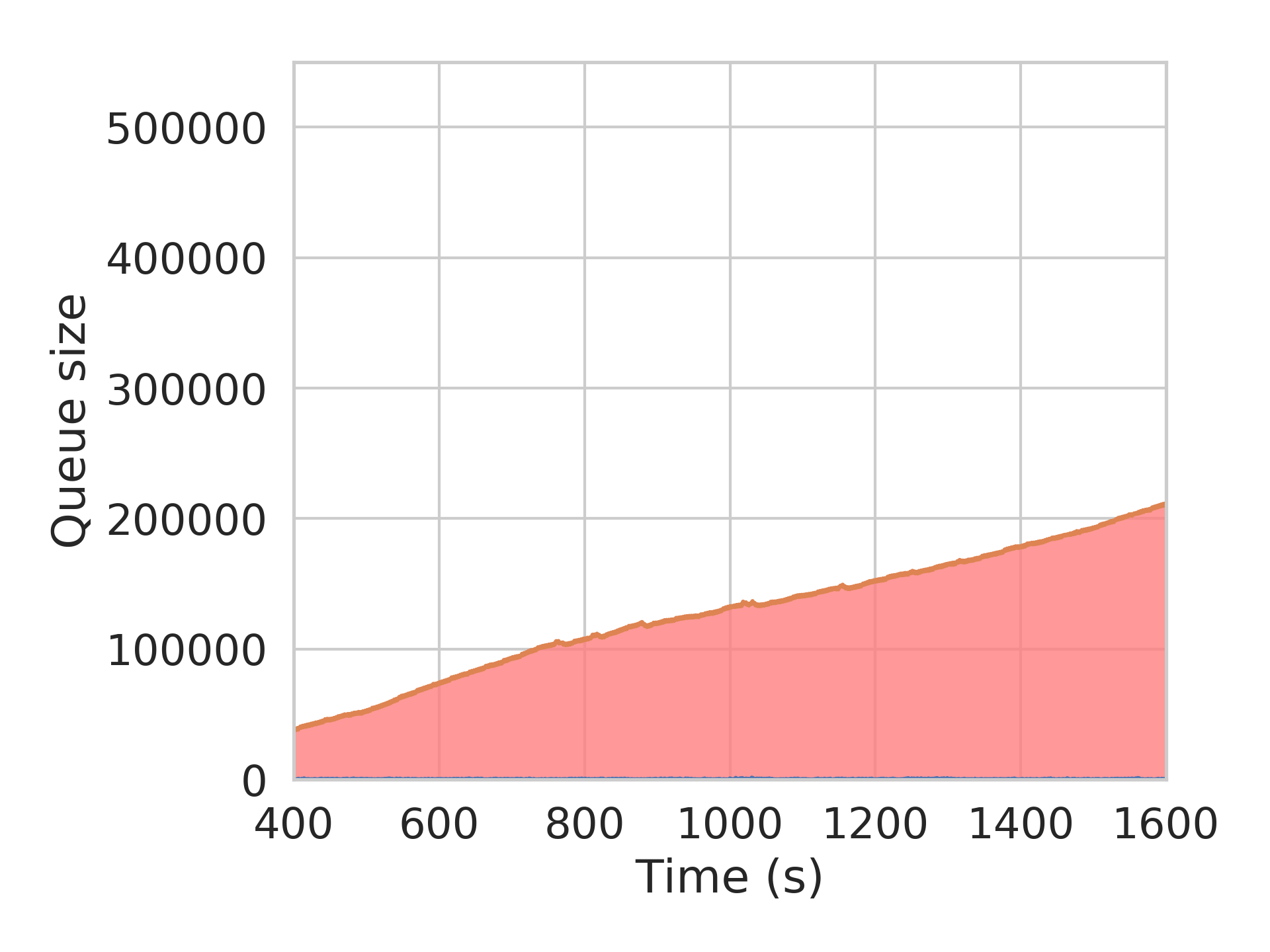}
 	\label{fig:greedy_queue}}
~
\hspace{-20px}
 \subfloat[Metis k-way]{
  	\includegraphics[width=0.25\textwidth]{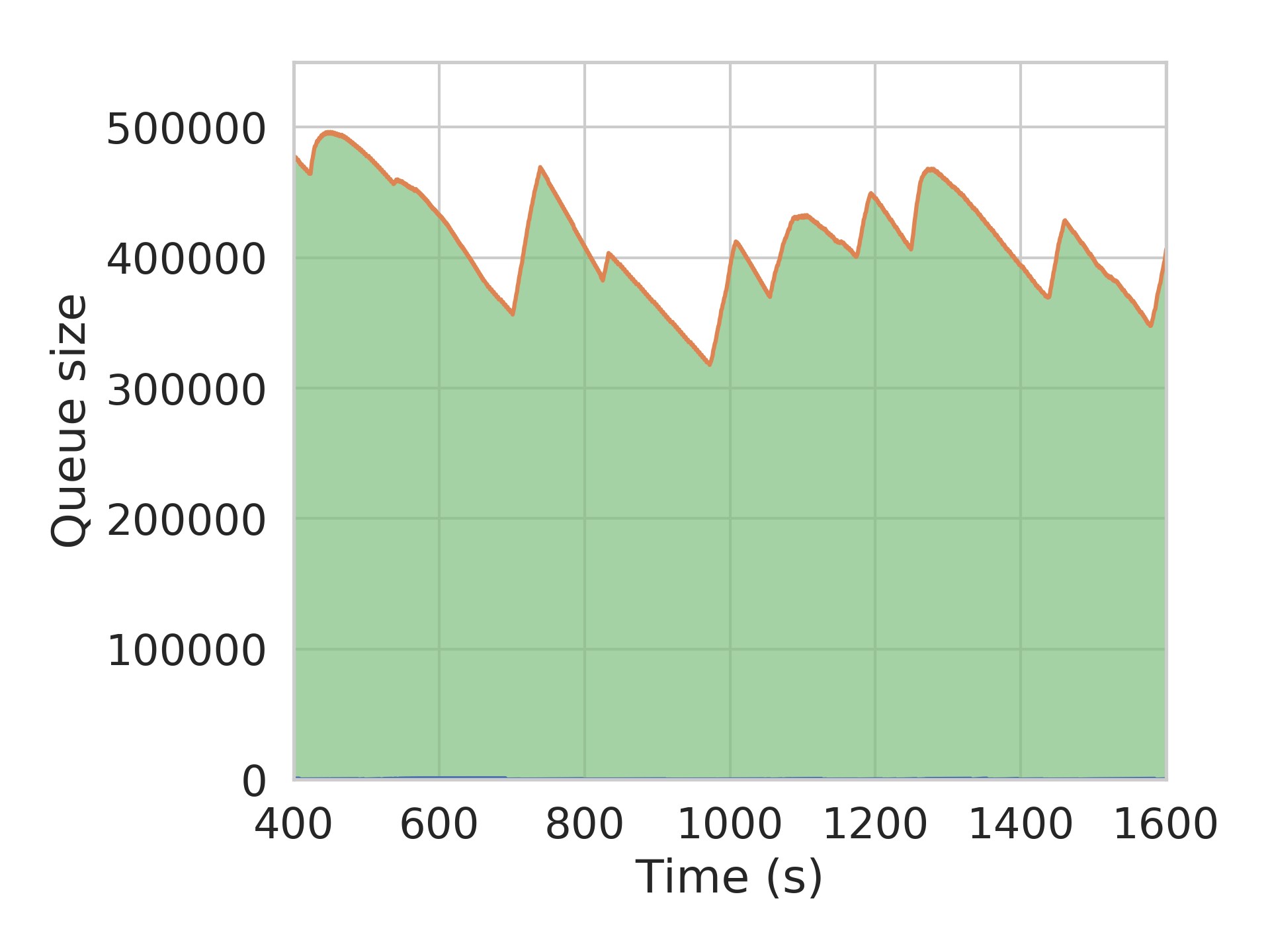}
  	\label{fig:metis_queue}}
    
    \caption{Maximum and minimum queue size of shards over time}
 	\label{fig:queue_size}
\end{figure}

To observe the congestion, we plot out the change on queue sizes of shards under four sharding methods as in Fig. \ref{fig:queue_size}. The congestion on Metis method is from the fact that Metis tends to put a large amount of consecutive transactions into one same shard. Thus, the pattern, in which some shards are overwhelmed while the others have no transactions, happens frequently in Metis. Greedy also met a situation that some shards have no transaction in several moments but in overall, transactions are splitted more equally than Metis among remaining shards. Such event does not happen in OptChain and OmniLedger. However, as OmniLedger is not capable of running with a transactions rate of 6000 tps, shards queue size will increase linearly overtime. OptChain performs the best in term of load balancing among shards. We can see that both the maximum and minimum queue size in OptChain are consistent and stable. Also, in worst case, a queue size in OptChain only reaches up to $\approx$ 44,000 transactions while these numbers in Metis, Greedy and OmniLedger are 507000, 230000 and 499000 respectively. 

\begin{figure}[h]
	\centering
	\includegraphics[width=0.7\linewidth]{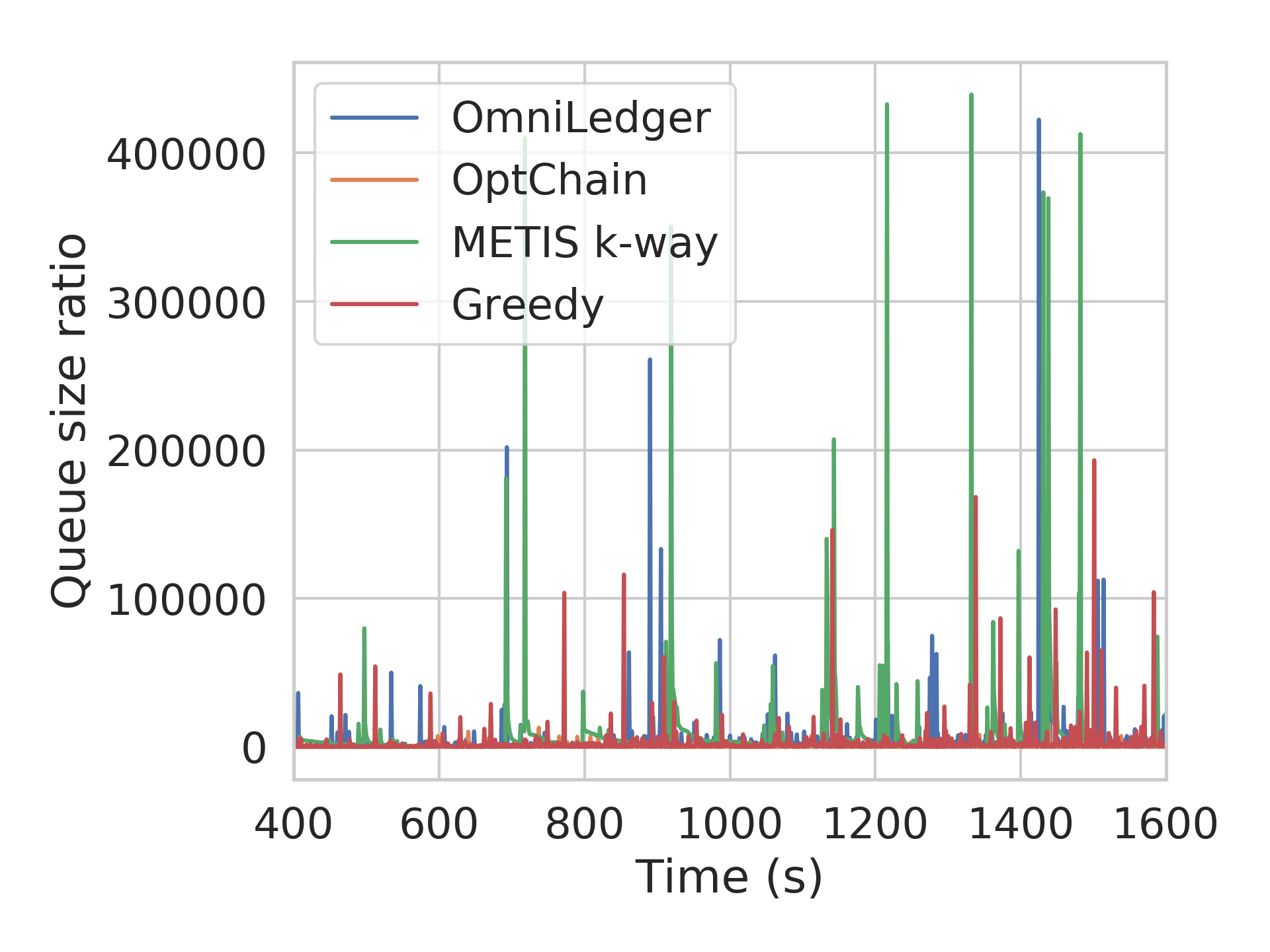}
	\caption{Queue size ratio}
	\label{fig:pr-q}
\end{figure}

To have a clearer observation of this behavior, we calculate the ratio between the maximum and minimum queue size of each algorithm at each simulation timestamp and compare them. \cref{fig:pr-q} illustrates this comparison where we can clearly see how inefficient the Metis and Greedy are in terms of temporal balance. Being unable to handle the load balancing among shards not only affects the throughput negatively but also notably increases the system's average latency as we shall see in the following experiments.

\subsubsection{Transaction Latency} Next, we compare the average (maximum) transaction latency, i.e., transaction \emph{confirmation time}, among the different sharding approaches. As demonstrated in Fig. \ref{fig:impact}, all four sharding methods share the same behavior: at a certain transaction rate, the average transaction latency decreases significantly when the number of shards increases. Therefore, all four methods performs their best in term of latency with the configuration of 16 shards as the number of transactions in each shard is low enough.

\begin{figure}
	\centering
	\hspace{-20px}
	\subfloat[ Number of shards =  16]{
	    \includegraphics[width=0.5\linewidth]{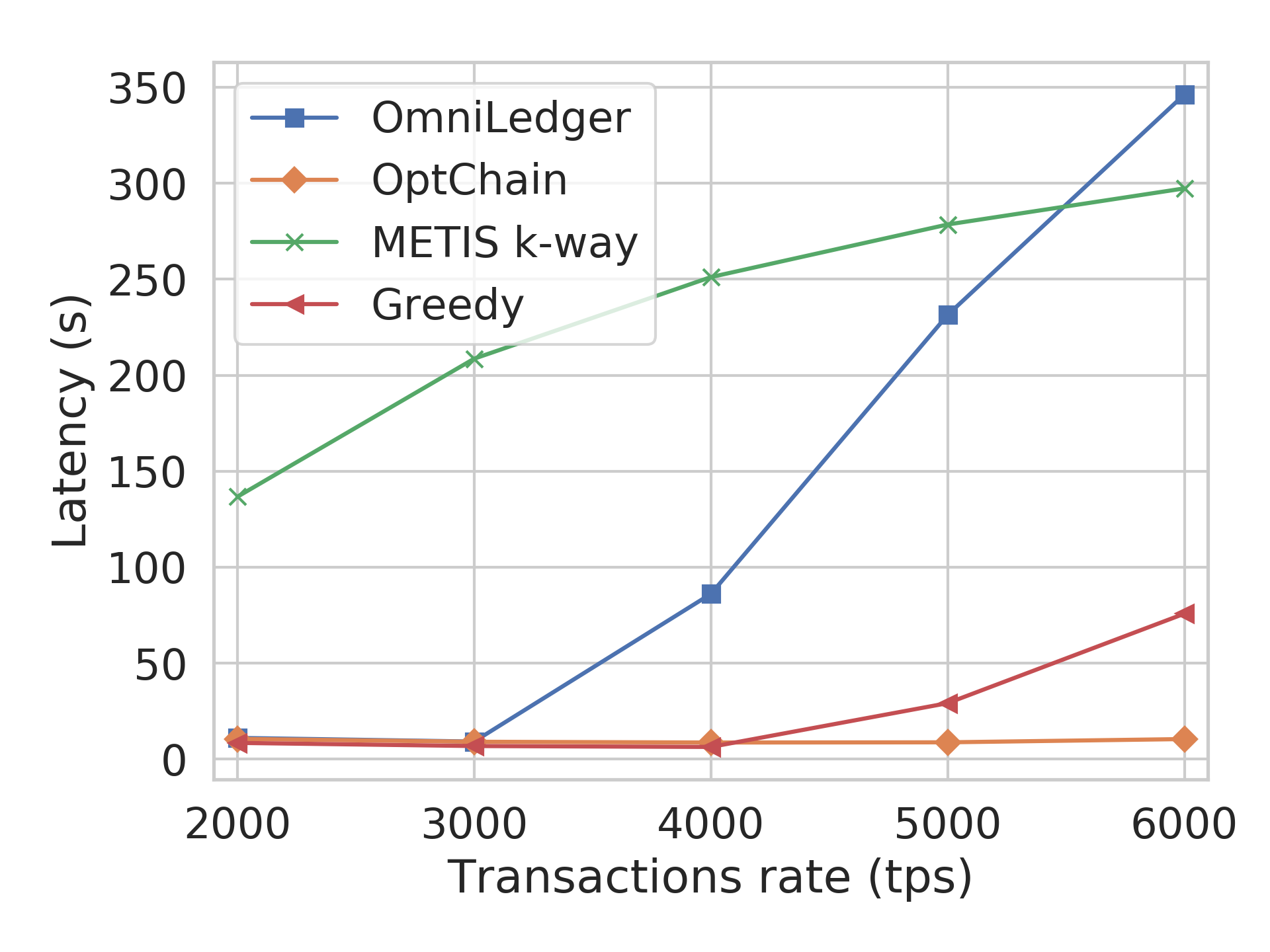}
	    \label{fig:lat-rate}
	}
	\subfloat[Vary transactions rate and \#shards]{
	    \includegraphics[width=0.5\linewidth]{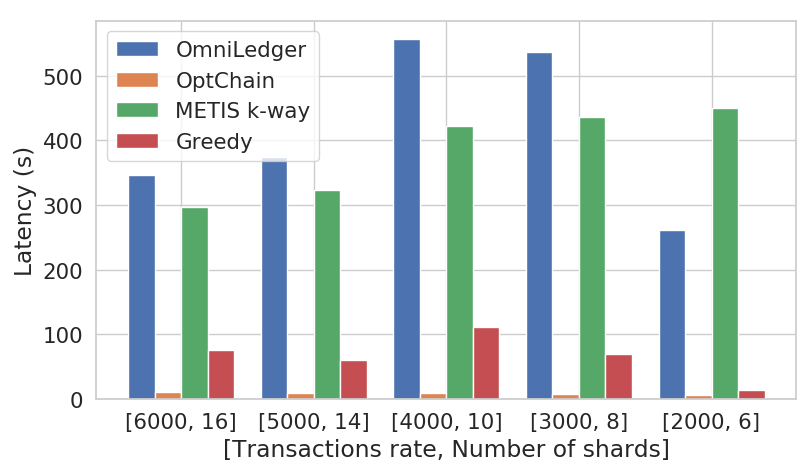}
	    \label{fig:lat-vary}
	}
	\caption{Average transaction latency}
	\label{fig:lat}
\end{figure}

To closely evaluate the latency, we varied the transaction rate while keeping the number of shards at 16 as shown in \cref{fig:lat-rate}. Clearly there exists a considerable gap between OptChain's average latency and other methods' results. OptChain always achieves the best performance comparing to the others. In fact, at the transactions rate of 4000 tps, the system only takes 8.7 seconds to process a transaction in average. At low transactions rate (i.e., 2000 to 3000 tps), we can see that all algorithms except for Metis have good latency in general. This is because the system can balance between throughput and transaction rate at these ranges, thus no backlogging happens. 

As we increase the rate to 6000 tps, our algorithm still maintains a good latency, while we can see a significant increase in other algorithms. In particular, at 6000 tps, we reduce up to 93\% the latency comparing to the OmniLedger. This comes from the fact that OmniLedger can not tolerate 6000 transactions per second in such settings. Thus, the queuing delay (time staying in queue) of a transaction will increase overtime and come to infinity. This behavior can also be observed in Greedy. As for Metis, even though it has the least amount of cross-shard transactions, it still gets really high average latency. This issue can be explained as Metis failed to achieve the temporal balance in queue size between shards as we mentioned above, causing there were only some active shards at a time and exacerbating the final average latency.


Next, we measure the system's latency with different combinations of transactions rate and number of shards as in \cref{fig:lat-vary}. Specifically, we set the configuration in the same way as \cref{fig:thru-shard}. Again as can be seen in \cref{fig:lat-vary}, OptChain obtains the best performance in term of balancing throughput and transaction rate. At these configurations, the highest average latency of OptChain is only 10.5 seconds at the transactions rate of 6000 tps and 16 shards, while OmniLedger reaches 346.2 seconds at this configuration. In addition, at the transactions rate of 4000 and 10 shards, OptChain reduces up to 98\% the latency as compared to OmniLedger.


\begin{figure}
    \hspace{-20px}
	\centering
	\subfloat[Number of shards = 16]{
	    \includegraphics[width=0.5\linewidth]{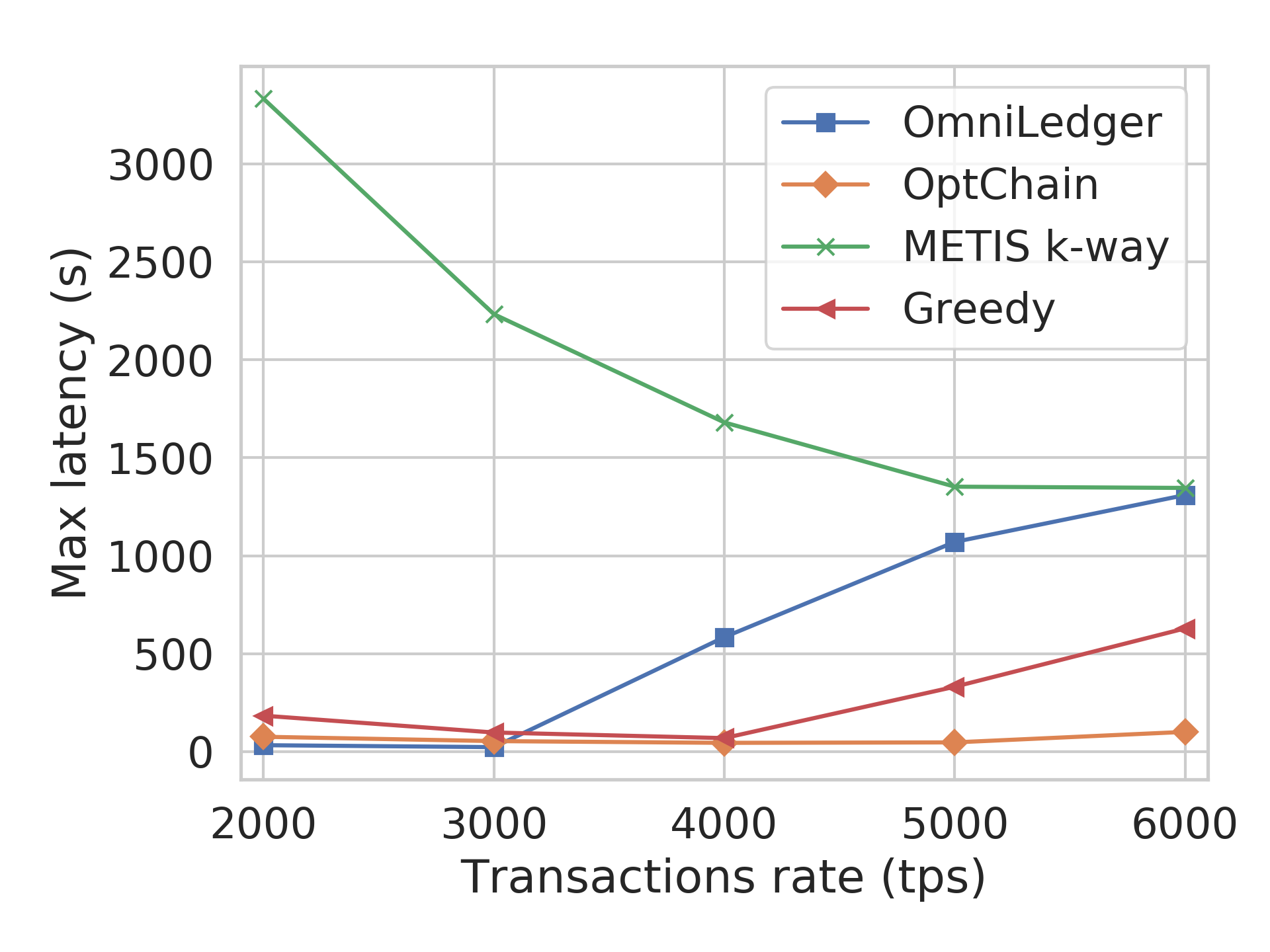}
	    \label{fig:maxlat-shard}
    }
	\subfloat[Vary transactions rate and \#shards]{
	    \includegraphics[width=0.5\linewidth]{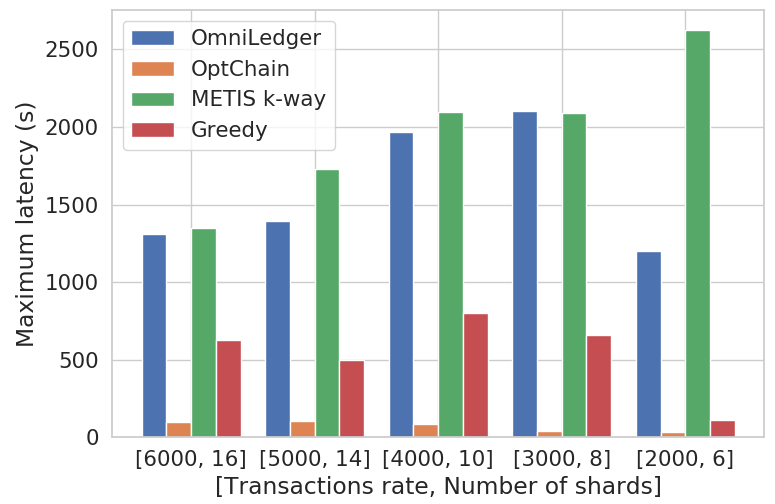}
	    \label{fig:maxlat-rate}
	}
	\caption{Maximum transaction latency}
	\label{fig:maxlat}
\end{figure}

\cref{fig:maxlat-shard} illustrates the maximum latency to process transactions with 16 shards at different transaction rates. At 6000 tps, OptChain takes at most 100.9 seconds for a transaction while OmniLedger, Metis, and Greedy respectively takes 1309.5, 1345.9, and 628.9 seconds. This result, together with \cref{fig:lat}, again confirms that our algorithm significantly reduces the transaction processing latency. Additionally, in \cref{fig:maxlat-rate}, we also plot the result of this metric at different combinations of transactions rate and number of shards as in the previous \cref{fig:lat-vary}. Among these configurations, the highest latency we ever reach is 102.7 seconds at the transactions rate of 5000 tps and 14 shards. At this point, OmniLedger, Metis, and Greedy respectively takes at most 1397.0, 1730.0, and 497.0 seconds to process one transaction.

\begin{figure}
	\centering
	\includegraphics[width=0.7\linewidth]{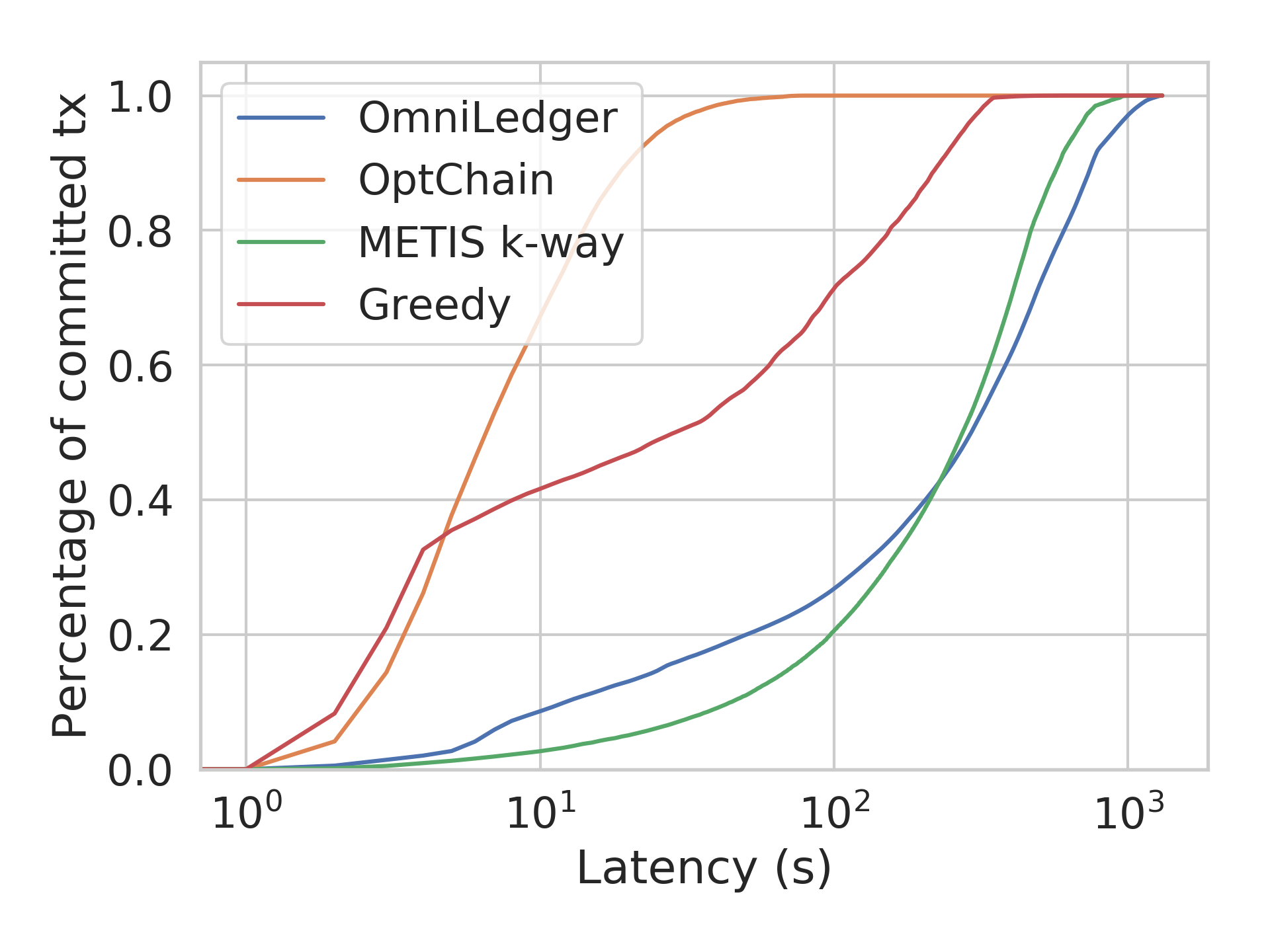}
	\caption{Latency distribution}
	\label{fig:lat-dist}
\end{figure}

For a more detailed view on the impact of algorithms on system's latency, in \cref{fig:lat-dist}, we present the cumulative distribution of the latency when we set the shards number to 16 and transactions rate 6000 tps. As can be seen, up to 70\% of the transactions are processed within 10 seconds with OptChain. For other algorithms, within this time frame, only 41.2\%, 7.9\%, and 2.4\% of the transactions were completed with Greedy, OmniLedger, and Metis, respectively. 

\subsection{Summary of results}
The experiments conducted in this section notably show that the proposed OptChain significantly outperforms other transaction sharding methods. To be specific, OptChain could reduce up to 93\% the latency and increase 50\% the throughput in comparison to OmniLedger. Furthermore, OptChain is more scalable than OmniLedger where it can scale up to the rate of 6000 transactions per second and 16 shards. Meanwhile, for a certain transaction rate, OptChain requires less shards than OmniLedgers to guarantee the system performance without backlogging. In overall, Optchain's good performance comes from the fact that OptChain is able to concurrently handle two main factors that hinder the system performance: (1) reducing \crsh s and (2) temporally balancing workload between shards.

\begin{figure}
	\centering
	\includegraphics[width=0.7\linewidth]{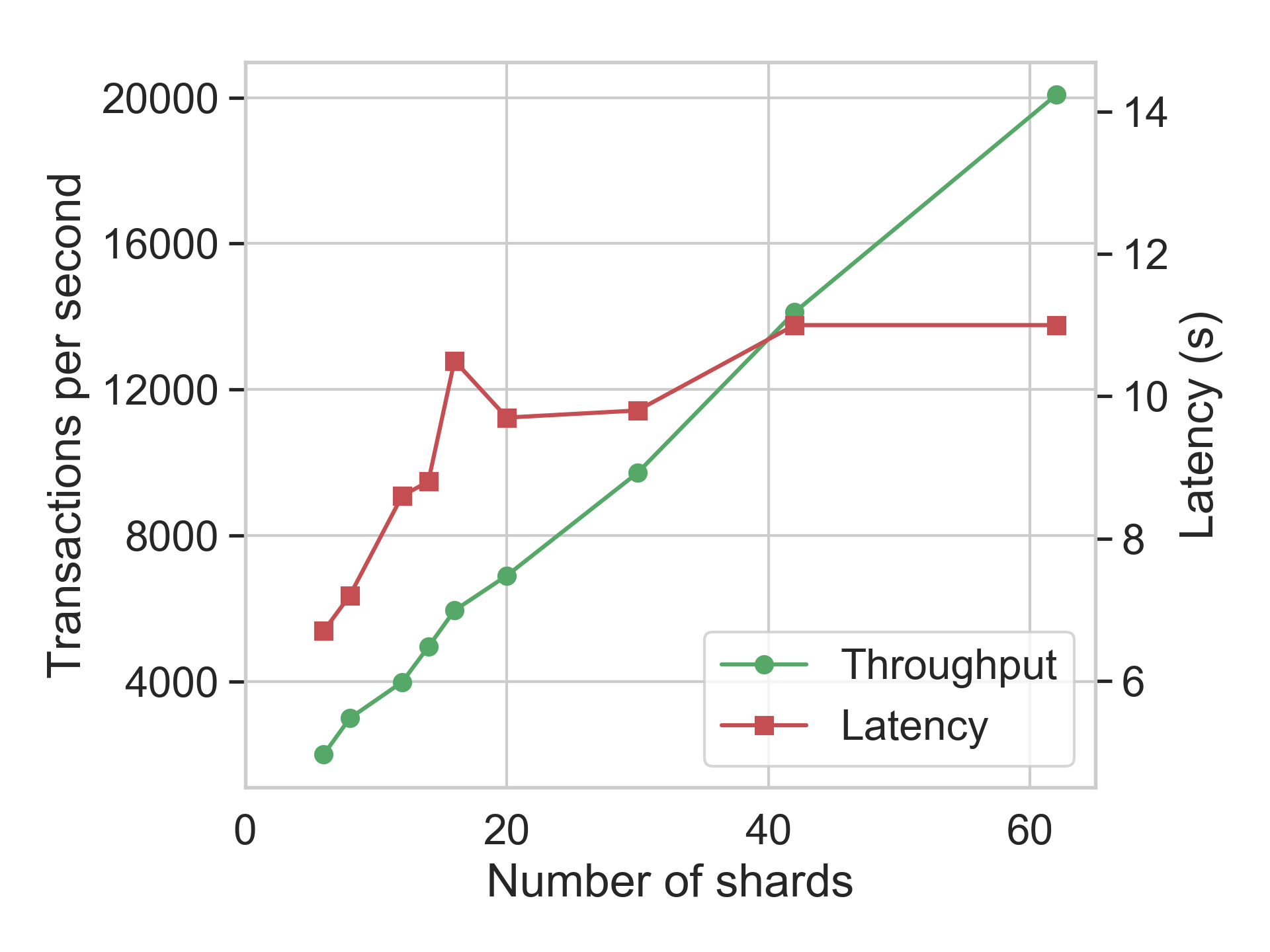}
	\caption{OptChain scalability}
	\label{fig:scale}
\end{figure}

The highest transaction rate \optch~can scale up to (throughput is equal to transaction rate) with multiple number of shards is plotted out as in Fig. \ref{fig:scale}. The best throughput of \optch~is almost linear with the number of shards and it can reach above 20,000 tps with 62 shards. More importantly, when the throughput is comfortable with transaction rate, \optch~guarantees that the confirmation delay is never more than 11 seconds.   

\section{Conclusion} \label{sec:conclusion}
Handling cross-shard transactions has been a major challenge in research on blockchain sharding as they are the main factor that limits the scalability of the system. In this paper, we have proposed OptChain, a scalable protocol to optimally place transactions into shards, which guarantee reducing number of cross-shard transaction as well as temporally balancing workload between shards. In experiments, OptChain reduces the latency by 93\% and increases the throughput by 50\% in comparison with the state-of-the-art sharding system, OmniLedger. Finally, our empirical evaluation demonstrates that OptChain scales smoothly with transactions rate up to 6,000 transactions per seconds with 16 shards and potentially handle much higher rate with more shards.



\bibliographystyle{IEEEtran}
\bibliography{bibliography}

\end{document}